\documentclass[fleqn,10pt]{wlscirep}
\usepackage[utf8]{inputenc}
\usepackage[T1]{fontenc}
\usepackage{float}
\usepackage{graphicx}
\usepackage{amssymb}
\usepackage{textcomp}
\usepackage{subcaption}
\usepackage{multirow}
\usepackage{soul}
\usepackage{color}

\newcommand{\beginsupplement}{%
	\setcounter{table}{0}
	\renewcommand{\thetable}{S\arabic{table}}%
	\setcounter{figure}{0}
	\renewcommand{\thefigure}{S\arabic{figure}}%
	
}

\title{Estimation of Regional Economic Development Indicator from Transportation Network Analytics\footnote{This is a pre-print of an article submitted to \textit{Scientific Reports}. The final authenticated version is available online at: \url{https://doi.org/10.1038/s41598-020-59505-2}}}

\author[1]{Bin Li}
\author[2,*]{Song Gao}
\author[2]{Yunlei Liang}
\author[2]{Yuhao Kang}
\author[2]{Timothy Prestby}
\author[2]{Yuqi Gao}
\author[1]{Runmou Xiao}
\affil[1]{School of Automobile, Chang’an University, Xi'an, Shaanxi, 710064, China}
\affil[2]{Geospatial Data Science Lab, Department of Geography, University of Wisconsin-Madison, WI, 53706, USA}

\affil[*]{Corresponding author email: song.gao@wisc.edu}

\begin{abstract}
With the booming economy in China, many researches have pointed out that the improvement of regional transportation infrastructure among other factors had an important effect on economic growth. Utilizing a large-scale dataset which includes 3.5 billion entry and exit records of vehicles along highways generated from toll collection systems, we attempt to establish the relevance of mid-distance land transport patterns to regional economic status through transportation network analyses. We apply standard measurements of complex networks to analyze the highway transportation networks. A set of traffic flow features are computed and correlated to the regional economic development indicator. The multi-linear regression models explain about 89$\%$ to 96$\%$ of the variation of cities' GDP across three provinces in China. We then fit gravity models using annual traffic volumes of cars, buses, and freight trucks between pairs of cities for each province separately as well as for the whole dataset. We find the temporal changes of distance-decay effects on spatial interactions between cities in transportation networks, which link to the economic development patterns of each province. We conclude that transportation big data reveal the status of regional economic development and contain valuable information of human mobility, production linkages, and logistics for regional management and planning. Our research offers insights into the investigation of regional economic development status using highway transportation big data. 
\end{abstract}
\begin{document}


\flushbottom
\maketitle
%
%
\thispagestyle{empty}


\section*{Introduction} \label{section:intro}
With the booming economy in China, many researches have pointed out that the improvement of regional transportation infrastructure, the mobility of labor and capital, and industry reform along with other socioeconomic factors play an important role on economic growth. Timely estimation of social and economic status of cities and regions has important implications for enterprise investment and government policy making. Traditional inference approaches to economic status mainly rely on official reports and census surveys, which usually take a long period and are labor intensive. With the rapid development of information, communication and technology (ICT), new data sources of human activities \cite{liu2016online} and vehicle movement flow \cite{gao2013understanding,wang2019migration,zhao2018empirical}, air transport flow  \cite{wang2011exploring,huang2017comparison}, financial flow \cite{zhen2019analyzing}, information flow \cite{wang2019regional}, communication flow \cite{gao2013discovering,chi2016uncovering,penguncovering,gao2017uncovering} , and others \cite{ma2019typeface} have become available for better understanding and monitoring the status of our socioeconomic environments \cite{liu2015social,lin2019measuring}. Liu et al. \cite{liu2016online} found that online social activity could reflect the macro economic status of 282 prefecture-level cities in China. Recently, Gao et al. \cite{gao2019computational} conducted a comprehensive review on data resources, computational tools, analytical methods, theoretical models, and applications in computational socioeconomics.

In the past decades, a wealth of works have been dedicated to studying the pattern of human mobility involving passenger transportation \cite{brockmann2006scaling, gonzalez2008understanding, song2010limits, simini2012universal, ren2014predicting, yan2013diversity, yan2017universal}. A comparatively smaller literature has been dedicated to the pattern of transportation activity embedded in goods movement \cite{anas2007regional,rahimi2008inland}. The scarcity of reliable data sources on freight transportation appears to be one of the challenges \cite{fu2013research}.
Early studies rely on traditional freight traffic surveys, which are typically enterprise questionnaire surveys to obtain information such as the traffic volume and speed in specific road sections. For example, Ogunsanya \cite{ogunsanya1982spatial} examined the field survey data collected from several freight terminals using questionnaires and freight delivery records of major warehouses in Lagos, Nigeria. With floating car technology being increasingly applied in transportation, large-scale global positioning system (GPS) tracking data of trucks have been used by researchers in exploring the spatial interaction patterns. In addition, several applications have been created in transportation planning through the prediction of road travel time, logistic demand, and analysis of vehicle operation characteristics. Comendador et al. \cite{comendador2012gps} presented urban freight distributions in two Spanish cities using GPS data, and compared the general mobility patterns of five groups of freight according to the type of goods including frequency and average distance.
Fu and Shi \cite{fu2013research} analyzed the spatiotemporal features of freight truck trips using GPS data that record the truck operation in real time.
Zanjani et al. \cite{zanjani2015estimation} estimated origin-destination truck flows based on truck GPS data, and the results indicate the value of GPS data in freight travel demand modeling.
Mrazovic et al. \cite{mrazovic2017understanding} explored the spatial and temporal patterns of trucks based on data on urban freight deliveries. 
Boarnet et al. \cite{boarnet2017urban} explored the freight flows in Los Angeles and provided evidence on employment being an important driver of freight activity.
With the burgeon of new data sources on transportation of goods, De Montis et al. \cite{deMontis2007structure} obtained a weighted network representation in which the vertices correspond to the Sardinian municipalities in Italy and the weighted edges correspond to the amount of commuting traffic among them.
They characterized the structure of human traffic (at the intercity level) and investigated its relation to the topological structure of cities (defined by the connectivity pattern among them).
Zhao et al. \cite{zhao2018empirical} used logistics data on origin-destination flows of goods in Hong Kong and analyzed the intra-urban freight movement based on the gravity model with a specific interest in uncovering potential trends in the distance decay effect, reflected by the parameter $\beta$, through multiple years of observation.
They took the population of sub-districts in the area of study as the values for nodal attractions and estimated the distance decay parameter $\beta$ by selecting among the candidates the parameter value that yielded the best goodness of fit. 
With the obtained estimates for $\beta$, they demonstrated that the estimated interaction flows coincide reasonably well with observed interaction flows, which argues for the effectiveness of the gravity model in determining spatial interactions of freight movements. Recently,  Ding et al. \cite{ding2019application} reviewed the applications of complex network theory in urban traffic studies. Our study differs from theirs in terms of the activity space (i.e. intra-urban vs. inter-urban). Accordingly, our study investigates regional connections and includes both intracity and intercity transportation of people and goods through highways. 

Apart from the studies using data on coach (vehicle) services, the analysis of airline services \cite{choi2006comparing,wang2011exploring,xiao2013reconstructing} and railway services \cite{masson2009can} have been catching the attention of researchers on the studies of spatial interactions. Barrat et al. \cite{barrat2004architecture} provided an early study of the worldwide airports network including the traffic flow and its correlation with the topological structure. A closely related field of study on transportation is dedicated to investigating the impact of transportation infrastructure on socioeconomic status such as economic growth or population growth which Beyzatlar et al. \cite{beyzatlar2014granger}, Iacono and Levinson \cite{iacono2016mutual} investigated. The rapid development of high-speed rails plays an important role in regional economic development. Zheng and Kahnconnect \cite{zheng2013china} demonstrated the effects of China’s bullet trains in facilitating labor market integration and mitigating the housing price and living costs in megacities. Jia et al. \cite{jia2017no} showed that the high-speed rail construction has a positive effect on regional economic growth in China. Cheng et al. \cite{cheng2015high} and Chen $\&$ Vickerman \cite{chen2017can}  investigated how the new development of high-speed rail infrastructure impacts on the economy structure of cities and regions in Europe. However, the endogenous problem of inverse causality of development potential on infrastructure investment has been a hurdle in producing a convincing conclusion on causality. Although the new development of high-speed rails reduces the transportation costs of people between large cities, there is a significant reduction in GDP after the high-speed rail upgrade in the counties located along the affected railway lines. The reduction was largely driven by the concurrent drop in fixed asset investments of those bypassed counties \cite{qin2017no}. Gao et al. \cite{gao2017collective} introduced the concept of inter-industry and inter-regional learnings of regional economic development. Using 25 years of economic data in China between 1990 and 2015, they addressed the endogeneity concerns by using the difference-in-differences analysis. Results showed that the high-speed rail development increased the industrial similarity of connected pairs of neighboring provinces \cite{gao2017collective}.  

In this research, we examined the regional economic development indicator measured by the gross domestic product (GDP) values of cities and the relationship between GDP and transportation activities of human and goods in three provinces of China. Over 3.5 billion records of vehicle entry and exit data were collected in highway toll stations (287 in Liaoning, 421 in Jiangsu, and 335 in Shaanxi, respectively) using toll collection systems and surveys in these three provinces between years 2014 and 2017. The raw station-level data were then aggregated and summarized to the city level as our main focus is to investigate the relationship between regional economic development and the transportation networks. More detailed data, feature construction, and method descriptions can be found in the "Methods" section. We conducted the multiple linear regression analysis using a set of traffic flow features with/without regularization techniques, and the modeling results explained about 89$\%$ to 96$\%$ of the variation of cities' GDP across three provinces. To further investigate the regional spatial interaction characteristics, we then constructed the highway transportation networks to fit gravity models using annual traffic volumes of cars, buses, and trucks between pairs of cities for each province separately as well as together for the whole dataset. We found the temporal changes of distance-decay effects on spatial interactions in transportation networks, which link to the regional economic development patterns in each province. In summary, the major contributions of this research are three-fold: (1) we present an analytical workflow to investigate the regional economic development status using highway transportation big data; (2) the analyses of highway transportation networks using the gravity model and the principle component analysis provide a good interpretation of the spatial structure of regional highway transportation development and the temporal economic changes; (3) the weighted network measures using the traffic flows correlate better with regional economy than that using the physical distance-based ones. 

\section*{Results} \label{section:results}

\subsection*{Estimation of GDP from traffic flows}
\begin{figure}[H]
	\centering
	\includegraphics[width=0.9\linewidth]{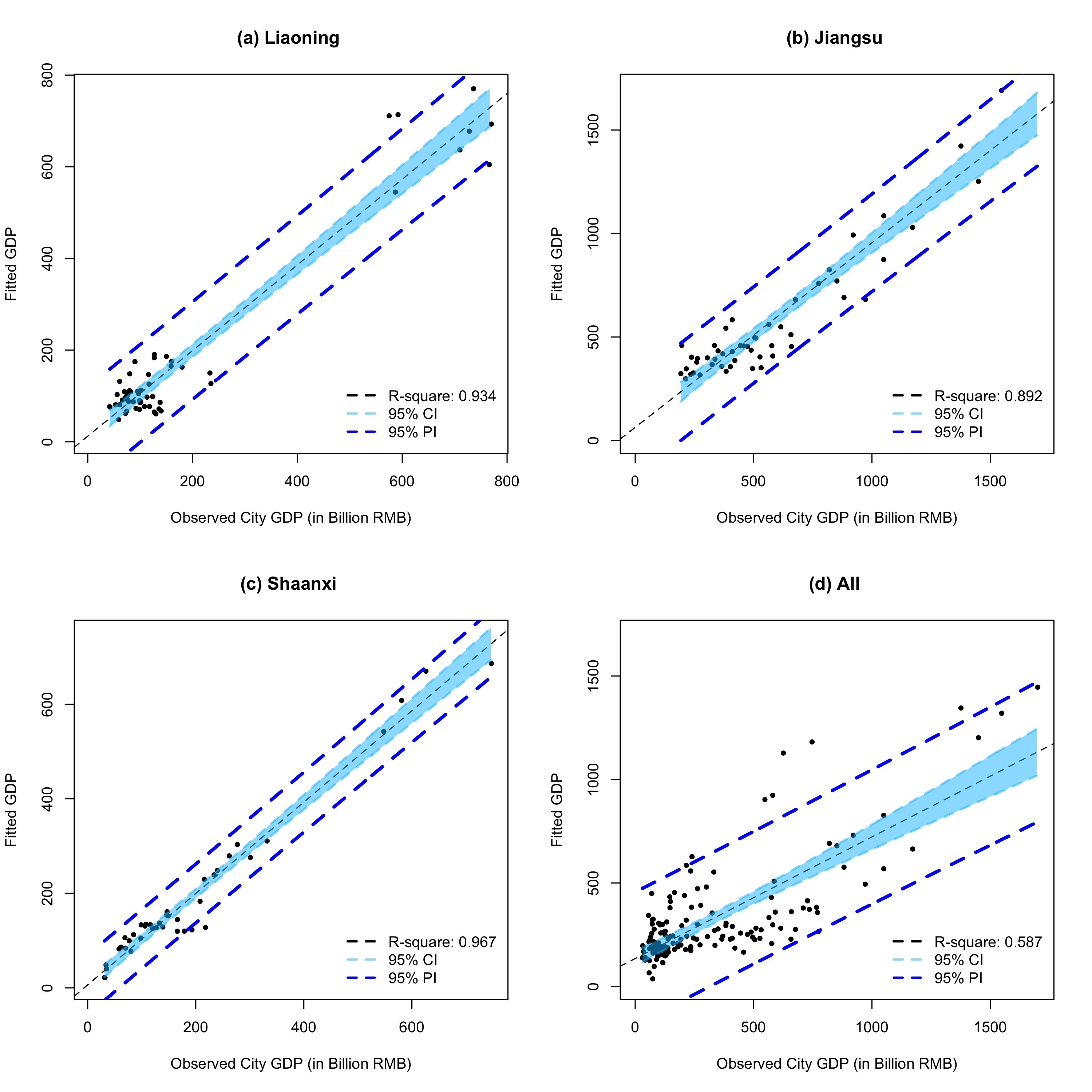}
	\caption{The relationships between the estimated and actual GDP values of cities in three provinces using multi-linear regression. (a) Liaoning province; (b) Jiangsu province; (c) Shaanxi province; (d) combine all cities together. (CI: confidence interval; PI: prediction interval.)}
	\label{fig:combine_LR_GDP}
\end{figure}

\begin{figure}[H]
	\centering
	\includegraphics[width=\linewidth]{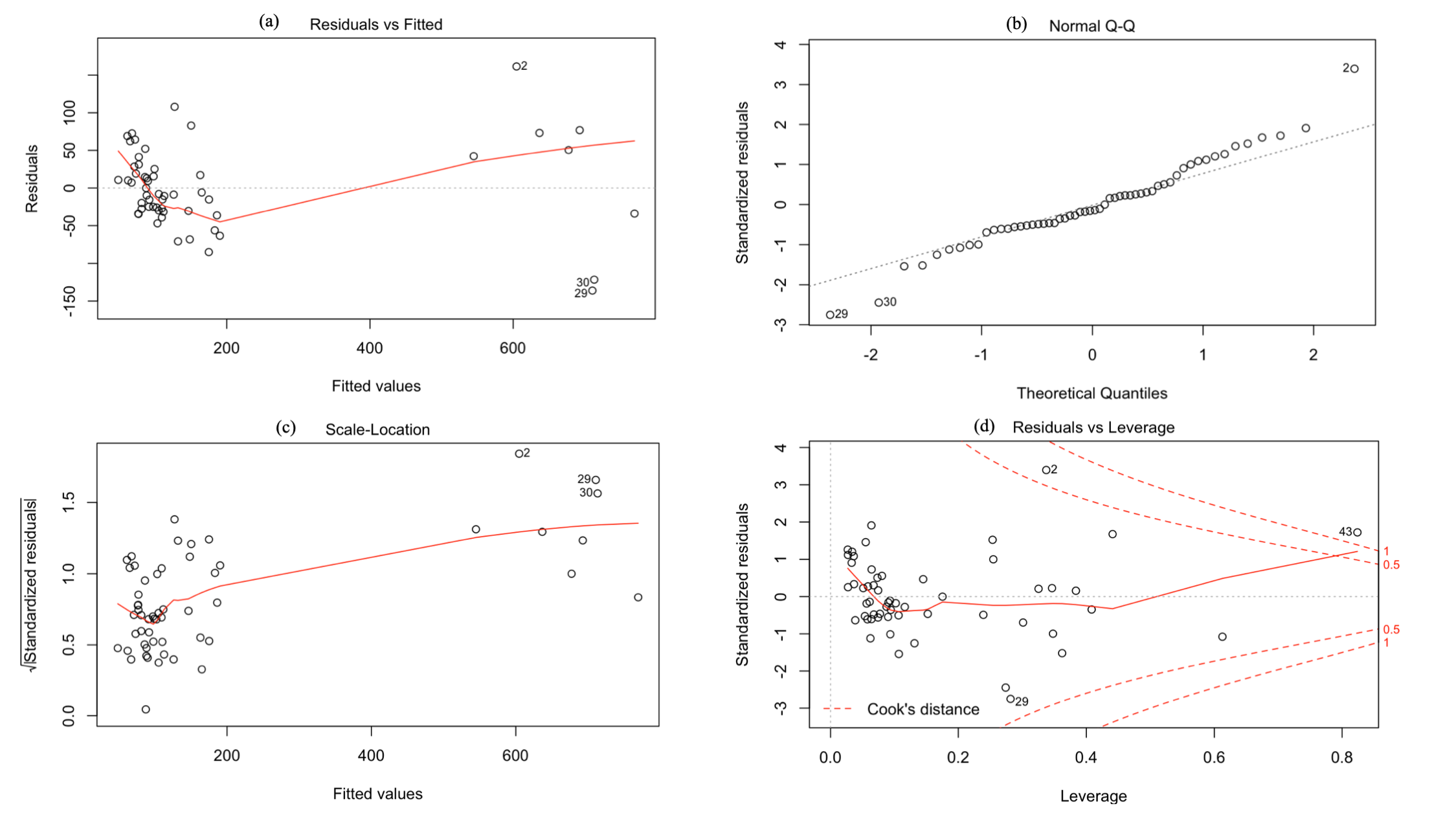}
	\caption{The residual plots of predicting city GDP values  in Liaoning province using the multiple linear regression. (a) the scatter plot of residuals and fitted values; (b) the Normal Quantile-Quantile plot for standardized residuals; (c) the Scale-Location plot to check the residual spread;  (d) the Residuals vs Leverage plot to find influential samples if any.}
	\label{fig:residualplots_LN}
\end{figure}

\begin{figure}[H]
	\centering
	\includegraphics[width=\linewidth]{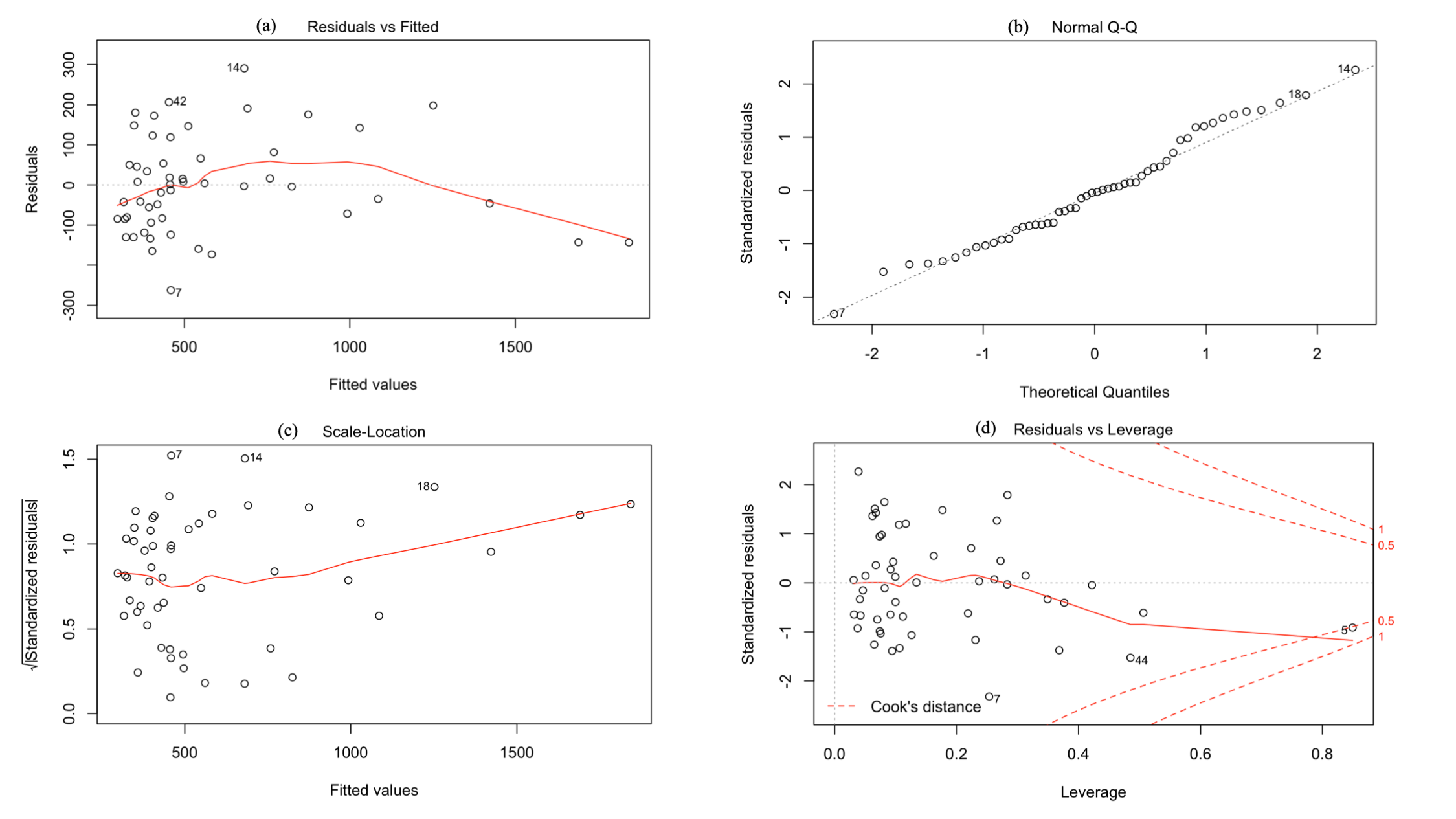}
	\caption{The residual plots of predicting city GDP values  in Jiangsu province using the multiple linear regression. (a) the scatter plot of residuals and fitted values; (b) the Normal Quantile-Quantile plot for standardized residuals; (c) the Scale-Location plot to check the residual spread;  (d) the Residuals vs Leverage plot to find influential samples if any.}
	\label{fig:residualplots_JS}
\end{figure}

\begin{figure}[H]
	\centering
	\includegraphics[width=\linewidth]{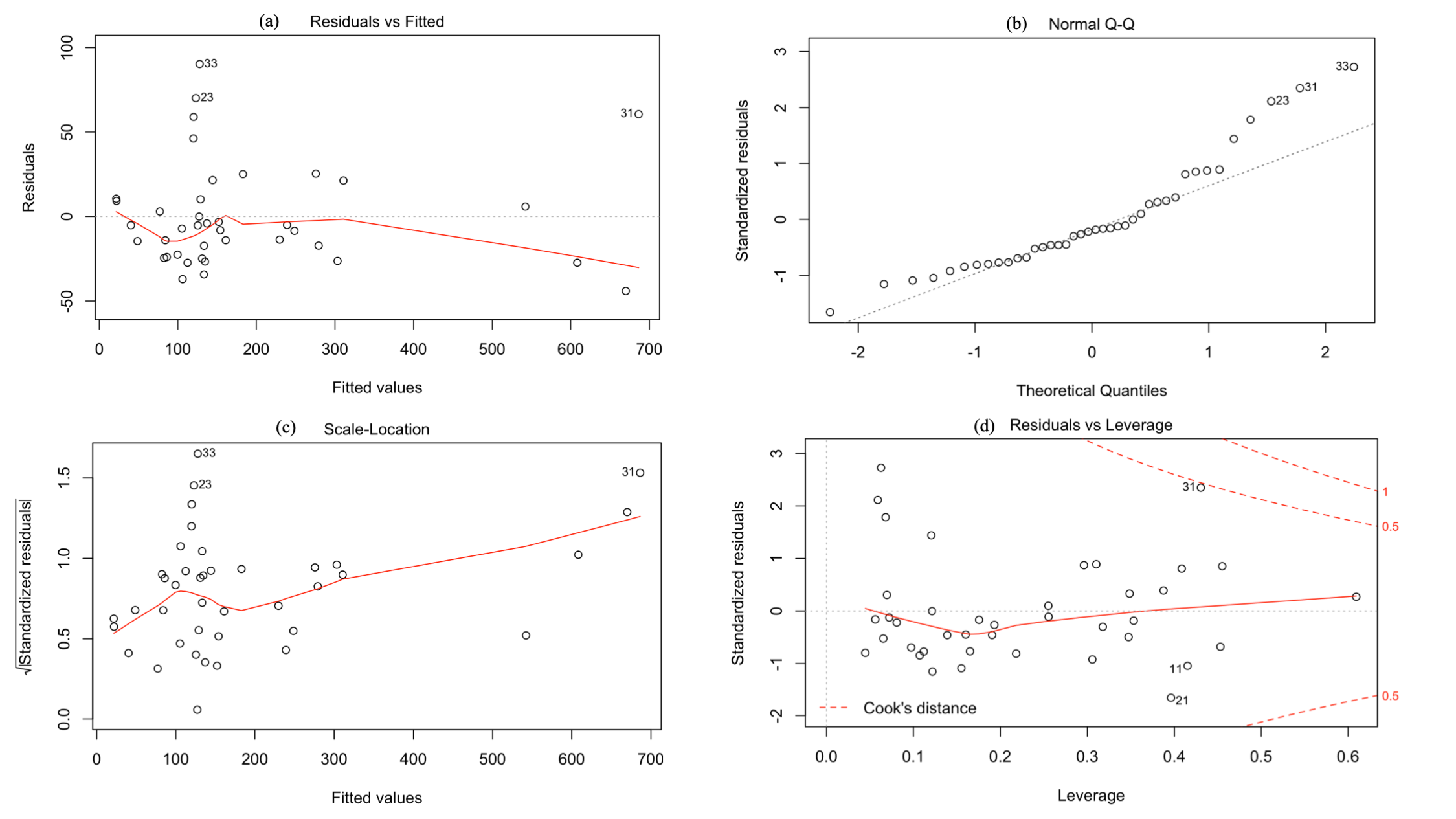}
	\caption{The residual plots of predicting city GDP values  in Shaanxi province using the multiple linear regression. (a) the scatter plot of residuals and fitted values; (b) the Normal Quantile-Quantile plot for standardized residuals; (c) the Scale-Location plot to check the residual spread;  (d) the Residuals vs Leverage plot to find influential samples if any.}
	\label{fig:residualplots_SX}
\end{figure}

\begin{figure}[H]
	\centering
	\includegraphics[width=\linewidth]{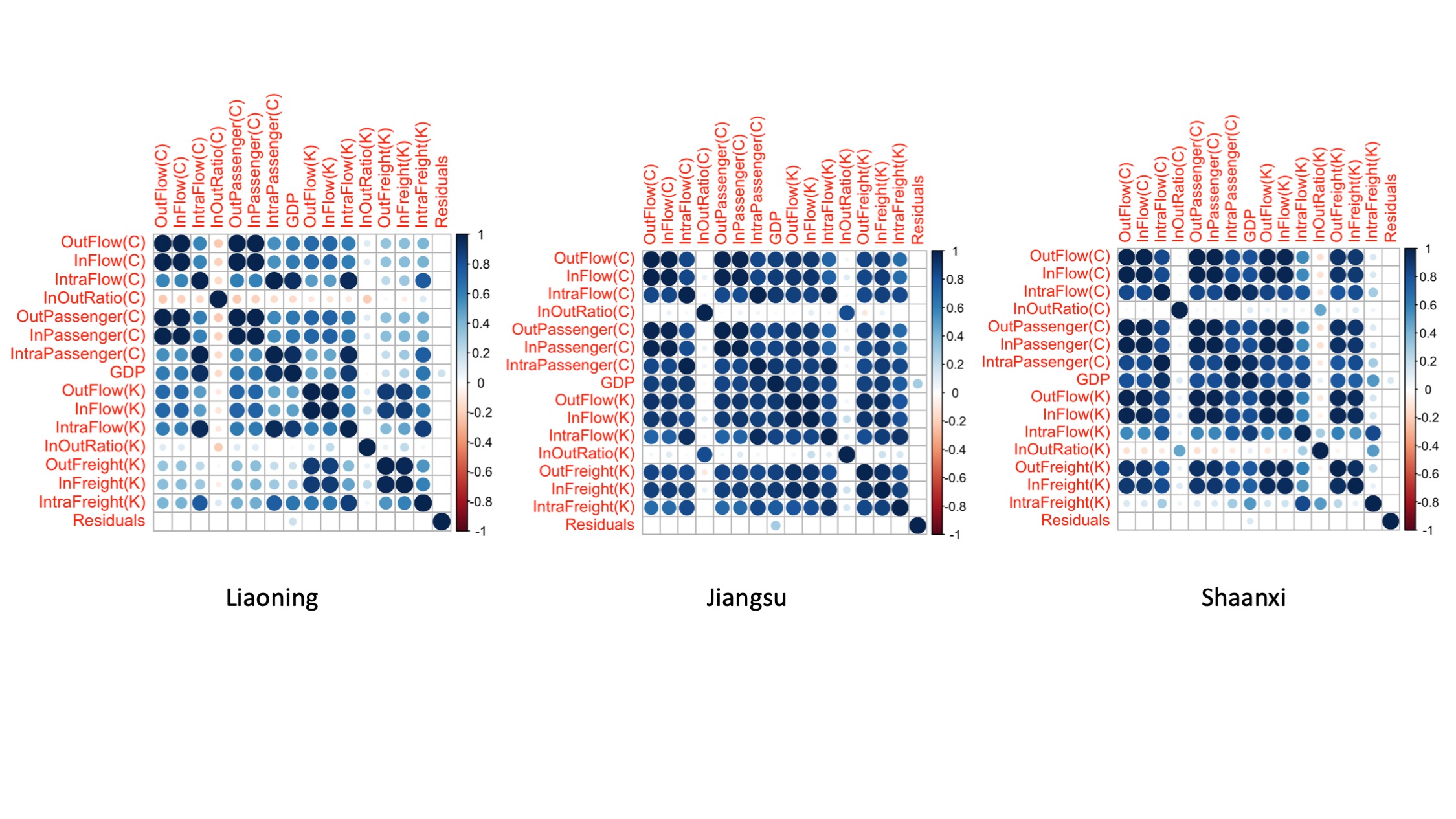}
	\caption{The visualization of correlation matrix of predictors and residuals in each province. }
	\label{fig:correlogram_allfeatures}
\end{figure}

The relationship between the fitted city GDP values using the multiple linear regression (MLR) models of transportation features and the actual city GDP in three provinces (i.e., Liaoning, Jiangsu, and Shaanxi) are summarized in Fig. \ref{fig:combine_LR_GDP}. It shows that simple transportation flow features (i.e., intra-city and inter-city flows of cars, buses, and trucks) extracted from the transportation networks of cities (in Equation 2) can explain the variation of the economic development indicator (i.e., GDP) among cities very well, with the goodness of fit: R-squared of 0.934 (in Liaoning province), 0.892 (in Jiangsu province), and 0.967 (in Shaanxi province). In addition, the R-squared further increased a margin (+0.025 in Liaoning, +0.021 in Jiangsu, and +0.014 in Shaanxi) by including the volume of passengers in cars $\&$ buses and the freight truck weights in the MLR model. After transforming the  dependent variable (GDP) with the natural log form (Ln), the generalized linear model had goodness of fit R-squared of 0.849 (in Liaoning), 0.727 (in Jiangsu), and 0.914 (in Shaanxi), respectively. The results are not as good as the original MLR approach. The prediction root-mean-square error (RMSE) of city GDP using original MLR model in three provinces are 53.5 (Liaoning), 119.8 (Jiangsu), and 30.11 (Shaanxi) billion CNY, respectively. As shown in the residual plots Figs. \ref{fig:residualplots_LN}, \ref{fig:residualplots_JS}, \ref{fig:residualplots_SX}, and \ref{fig:correlogram_allfeatures}, the residuals center on zero and are not correlated with any predictors, which indicate that these models' predictions have a relatively constant variance and show homoscedasticity and normality. In addition, by applying two regularized regression methods: the Ridge regression \cite{hoerl1970ridge} and the least absolute shrinkage and selection operator (LASSO) regression \cite{tibshirani1996regression,dong2019predicting}, the smallest RMSE values were 54.2 (Liaoning), 121.0 (Jiangsu), and 30.66 (Shaanxi) billion CNY, respectively. Moreover, these two methods only select a few predictors while reducing the coefficients of other highly correlated predictors to zero. It helps mitigate the problem of multicollinearity in MLR  based on the ordinary least squares estimation, and the model still keeps a high goodness of fit for each province' data (See the 'Discussion and Conclusion' section in detail). However, the selected three provinces had different economic development patterns in the study period. Accordingly, the goodness of fit (R-squared = 0.587) of a universal MLR model which combines all the GDP values and traffic flow data across three provinces is not as good as that in each province. The corresponding prediction RMSE increased to about 212 billion CNY using MLR. This demonstrates the complexity and heterogeneity of the economic development in different cities and provinces. The findings correspond to existing research on the regional economic complexity in China using non-monetary metrics \cite{hidalgo2009building,gao2018quantifying}.  As for the impact of predictors, we found both similarities and variations of their standardized regression coefficients in different provinces (as shown in Table \ref{table:coefficients}). Specifically, the intracity flow of cars and buses has the largest positive impact to the city GDP with a standardized coefficient (0.362) in Jiangsu province, followed by the intercity truck in-flow (0.263) and the intercity car/bus in-flow (0.262), while in the case of Liaoning province, the  intercity car/bus in-flow has the largest positive standardized coefficient (7.371) followed by the intracity flow of cars and buses (1.134). In Shaanxi province, the intercity out-flow of cars and buses has the largest positive impact to the city GDP with a standardized coefficient (2.527), followed by the intracity flow of trucks (0.476) and the intracity flow of cars and buses (0.401).  With the Ridge and LASSO regressions, we found very similar results (in Table \ref{table:regularization}) to the non-regularized MLR regarding the predictor selection. Accordingly, car/bus intracity flow and intercity in-flow are both selected to explain the GDP variation among cities in Jiangsu and Liaoning provinces. One different feature selection result is that the intercity out-flow of trucks selected in the Ridge and LASSO regression models for determining cities' GDP in Jiangsu province had a smaller standardized coefficient than that of the intercity in-flow of trucks in the non-regularized MLR. As for the Shaanxi province, six out of eight transport predictors are selected and the intracity flows of cars $\&$ buses as well as trucks play an important role in predicting the city GDP as reported in Table \ref{table:regularization}. 

\begin{figure}[H]
	\centering
	\includegraphics[width=0.9\linewidth]{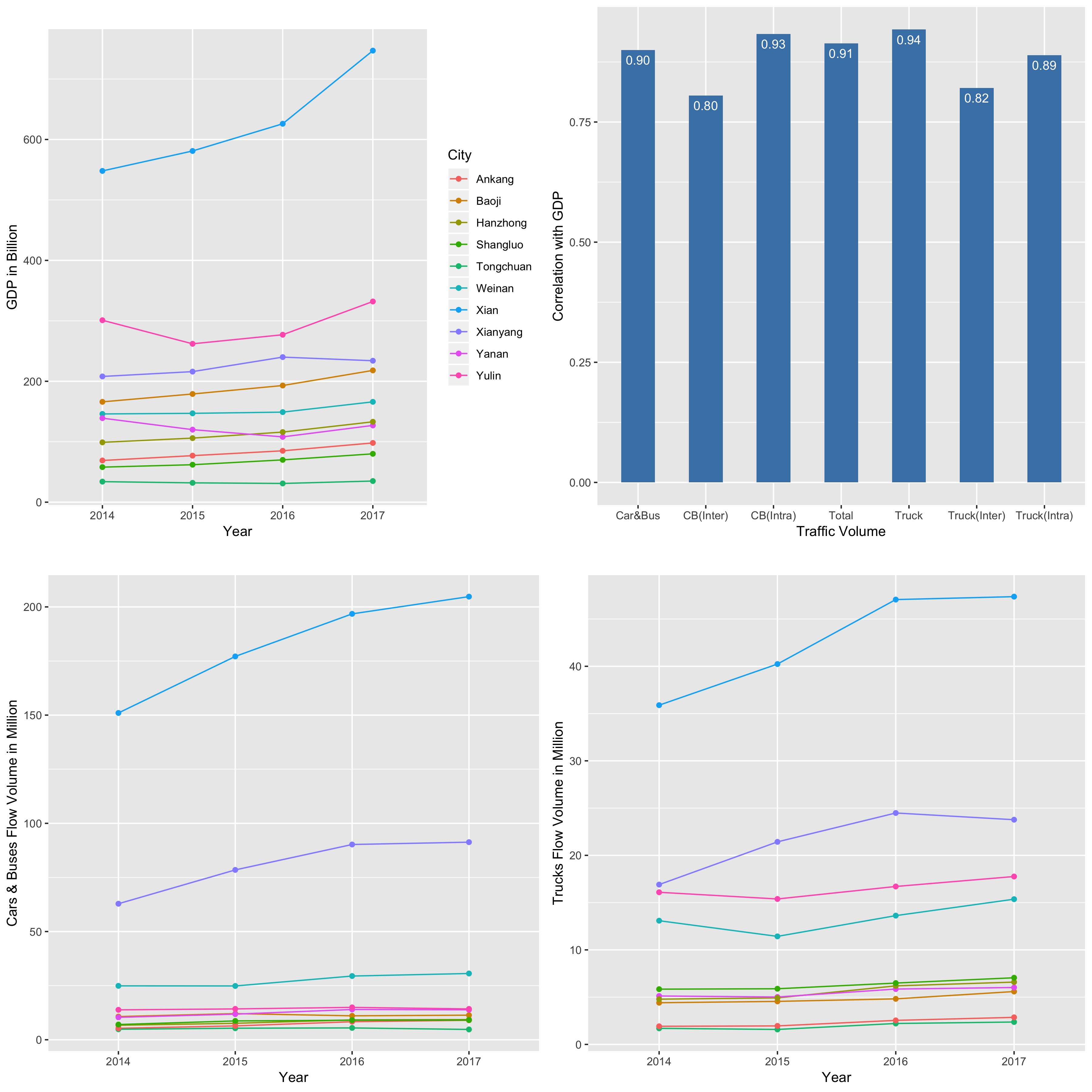}
	\caption{(a) The temporal changes of cities' GDP; (b) the correlation between city GDP and traffic volumes; (c) the temporal changes of traffic volumes of cars and buses;  (d) the temporal changes of traffic volumes of trucks in Shaanxi province.}
	\label{fig:GDP_transportation}
\end{figure}

\begin{figure}[H]
	\centering
	\includegraphics[width=\linewidth]{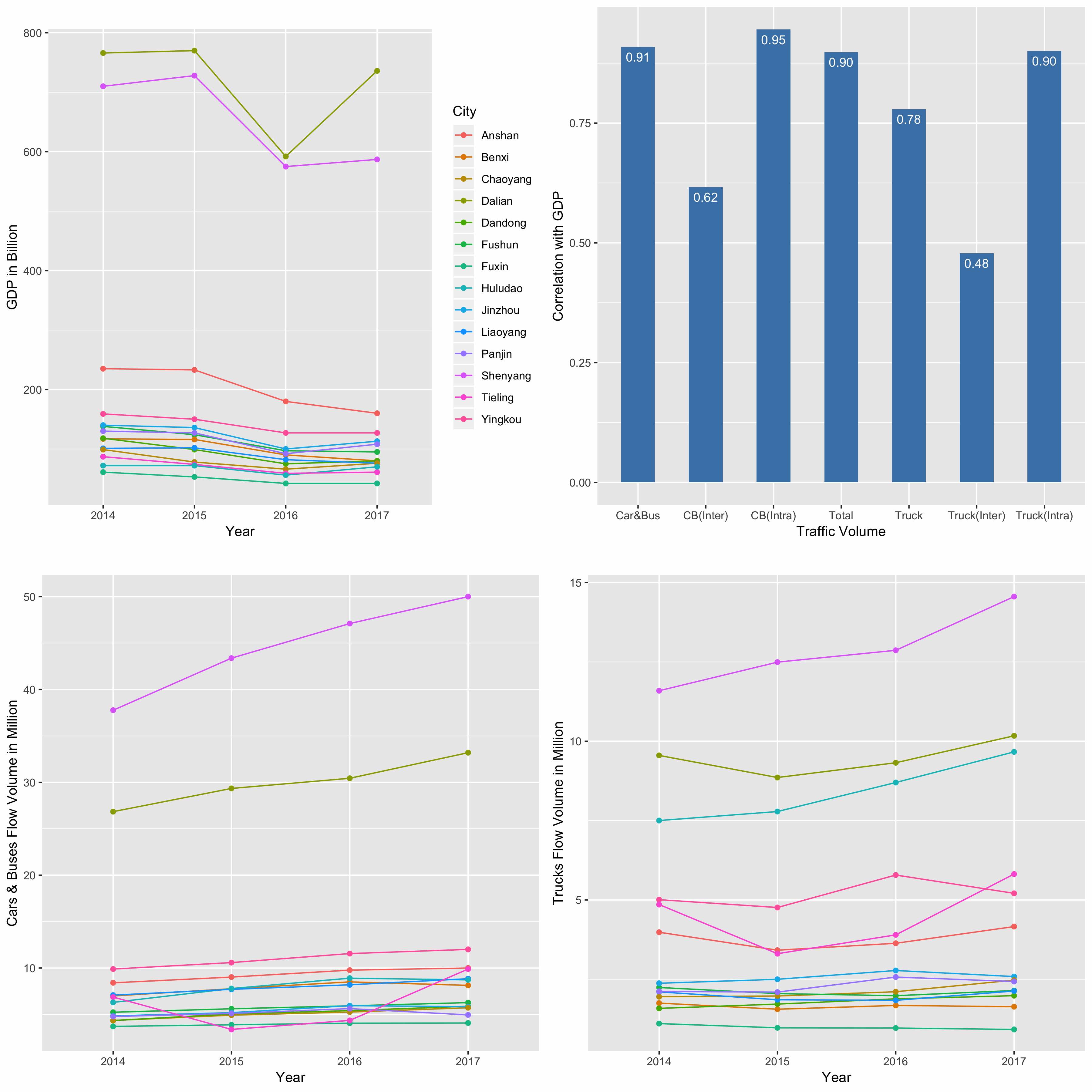}
	\caption{(a) The temporal changes of cities' GDP values; (b) the correlation between city GDP and traffic volumes; (c) the temporal changes of traffic volumes of cars and buses;  (d) the temporal changes of traffic volumes of trucks in Liaoning province.}
	\label{fig:LN_GDP_transportation}
\end{figure}

\begin{figure}[H]
	\centering
	\includegraphics[width=\linewidth]{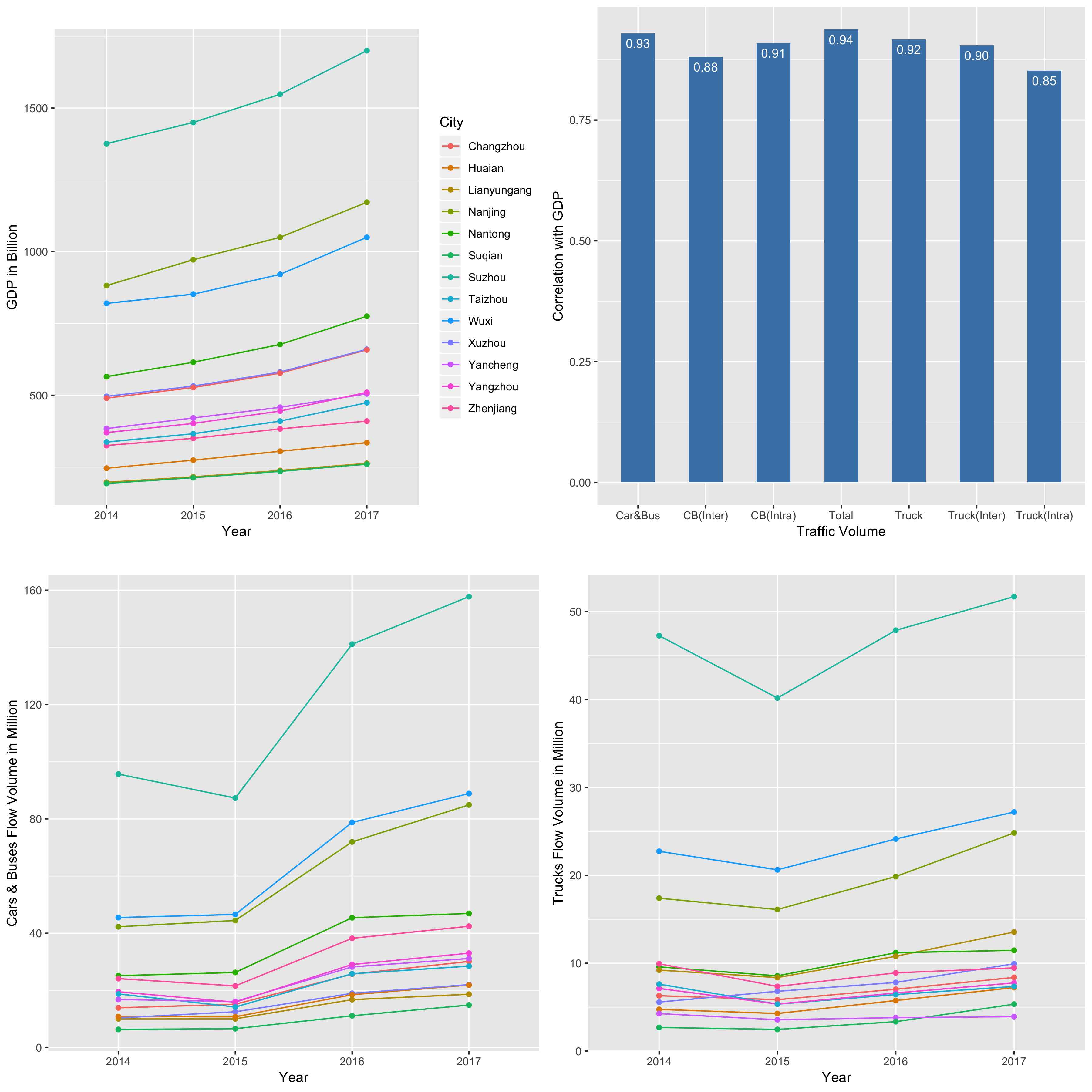}
	\caption{(a) The temporal changes of cities' GDP values; (b) the correlation between city GDP and traffic volumes; (c) the temporal changes of traffic volumes of cars and buses;  (d) the temporal changes of traffic volumes of trucks in Jiangsu province.}
	\label{fig:JS_GDP_transportation}
\end{figure}

In addition, Fig. \ref{fig:GDP_transportation} and supplementary figures 	\ref{fig:LN_GDP_transportation} and \ref{fig:JS_GDP_transportation}  show that the temporal changes of traffic volumes of cars \& buses and trucks over the consecutive four-year period match the city GDP overall changes well. Moreover, our experiments based on the total traffic volume (including all flows of cars \& buses and trucks) of each city over four years and city GDP data using a Bayesian structural time-series model \cite{brodersen2015inferring} verify that transportation infrastructure can boost economic growth (with p-value 0.002), since it has significant impact on transport activities including both human movements and freight flows.

\begin{table}[H]
				\centering
	\begin{tabular}{|c|c|c|c|}
		\hline
		Standardized Coefficients & Liaoning & Jiangsu & Shaanxi \\ \hline
		$I_C$ & 7.371    & 0.262   & -2.183  \\ \hline
		$O_C$ & -7.273   & -0.070  & 2.527   \\ \hline
		$N_C$ & 1.134    & 0.362   & 0.401   \\ \hline
		$R_C$ & -0.173   & 0.117   & 0.139   \\ \hline
		$I_K$ & -0.310   & 0.263   & 0.072   \\ \hline
		$O_K$ & 0.248    & 0.113   & -0.226  \\ \hline
		$N_K$ & -0.222   & 0.063   & 0.476   \\ \hline
		$R_K$ & 0.026    & -0.183  & -0.044  \\ \hline
	\end{tabular}
	\caption{The standardized multi-linear regression coefficients using the ordinary least squares technique. Note: intercity incoming flow of cars $\&$ buses ($I_{C}$), intercity outgoing flow of cars $\&$ buses ($O_{C}$), intracity flow of cars $\&$ buses ($N_{C}$), and the ratio of incoming/outgoing intercity flow for cars $\&$ buses ($R_{C}$), intercity incoming flow of trucks ($I_{K}$), intercity outgoing flow of trucks ($O_{K}$), intracity flow of trucks ($N_{K}$), and the ratio of incoming/outgoing intercity flow for trucks ($R_{K}$). }
	\label{table:coefficients}
\end{table}

\begin{table}[H]
	\centering
	\scalebox{0.85}{
		\begin{tabular}{|c|c|c|c|c|c|c|c|c|}
			\hline
			& Ridge $\lambda$ & Ridge R-squared & Ridge RMSE  & \begin{tabular}[c]{@{}c@{}}Ridge Selected Features \\ and Coefficients\end{tabular}      & LASSO $\lambda$ & R-squared & RMSE  & \begin{tabular}[c]{@{}c@{}}LASSO Selected Features \\ and Coefficients\end{tabular}                                                                                      \\ \hline
			Liaoning & 21.62                & 0.887              & 69.9          & \begin{tabular}[c]{@{}c@{}}Intercept: 4.97e+01\\ $I_C$: 1.18e-05\\ $N_{C}$:1.85e-05\end{tabular}                                & 0.06                 & 0.932              & 54.2          & \begin{tabular}[c]{@{}c@{}}Intercept: 6.64e+02\\ $I_{C}$: 4.4e-04\\ $O_{C}$: -4.2e-04\\ $N_{C}$: 3.30e-05\\ $R_{C}$: -6.5e+02\\ $I_{K}$: -2.45e-06\\ $O_{K}$: -2.0e-06\\ $N_{K}$: -3.2e-05\\ $R_{K}$: 5.26\end{tabular} \\ \hline
			Jiangsu  & 52.38                & 0.888              & 121.0         & \begin{tabular}[c]{@{}c@{}}Intercept: 2.9e+02\\ $I_{C}$: 3.2e-07\\ $O_{C}$: 6.8e-07\\ $N_{C}$: 1.0e-05\\ $O_{K}$: 3.4e-05\end{tabular}         & 18.39                & 0.882              & 124.2         & \begin{tabular}[c]{@{}c@{}}Intercept: 2.32e+02\\ $I_{C}$: 5.8e-06\\ $N_{C}$: 1.0e-05\\ $O_{K}$:3.8e-05\end{tabular}                                                                             \\ \hline
			Shaanxi  & 16.89                & 0.961              & 32.59         & \begin{tabular}[c]{@{}c@{}}Intercept: 7.92e+01\\ $I_{C}$: 6.61e-08\\ $O_{C}$: 1.29e-06\\ $N_{C}$: 2.82e-06\\ $N_{K}$: 1.27e-05\end{tabular} & 0.70                 & 0.965              & 30.66         & \begin{tabular}[c]{@{}c@{}}Intercept: -2.02e+02\\ $I_{C}$: 1.15e-06\\ $O_{C}$: 7.55e-07\\ $N_{C}$: 2.83e-06\\ $R_{C}$: 2.94e+02\\ $N_{K}$: 1.47e-05\\ $R_{K}$: -3.05e+01\end{tabular}                         \\ \hline
			All      & 25.23                & 0.570              & 213.4         & \begin{tabular}[c]{@{}c@{}}Intercept: 1.31e+02\\ $I_{K}$: 3.53e-05\\ $O_{K}$: 5.05e-06\\ $N_{K}$: 2.31e-05\end{tabular}                & 6.70                 & 0.569              & 213.5         & \begin{tabular}[c]{@{}c@{}}Intercept: 3.15e+02\\ $I_{K}$: 4.28e-05\\ $N_{K}$: 2.62e-05\\ $R_{K}$: -1.99e+02\end{tabular}                                                                       \\ \hline
	\end{tabular}}
	\caption{Ridge and LASSO regression analysis results. Note: intercity incoming flow of cars $\&$ buses ($I_{C}$), intercity outgoing flow of cars $\&$ buses ($O_{C}$), intracity flow of cars $\&$ buses ($N_{C}$), and the ratio of incoming/outgoing intercity flow for cars $\&$ buses ($R_{C}$), intercity incoming flow of trucks ($I_{K}$), intercity outgoing flow of trucks ($O_{K}$), intracity flow of trucks ($N_{K}$), and the ratio of incoming/outgoing intercity flow for trucks ($R_{K}$). }
	\label{table:regularization}
\end{table}

\subsection*{City attractiveness and  distance-decay effects }
\begin{figure}[H]
	\begin{subfigure}{0.45\linewidth}
		\centering
		\includegraphics[width=\linewidth]{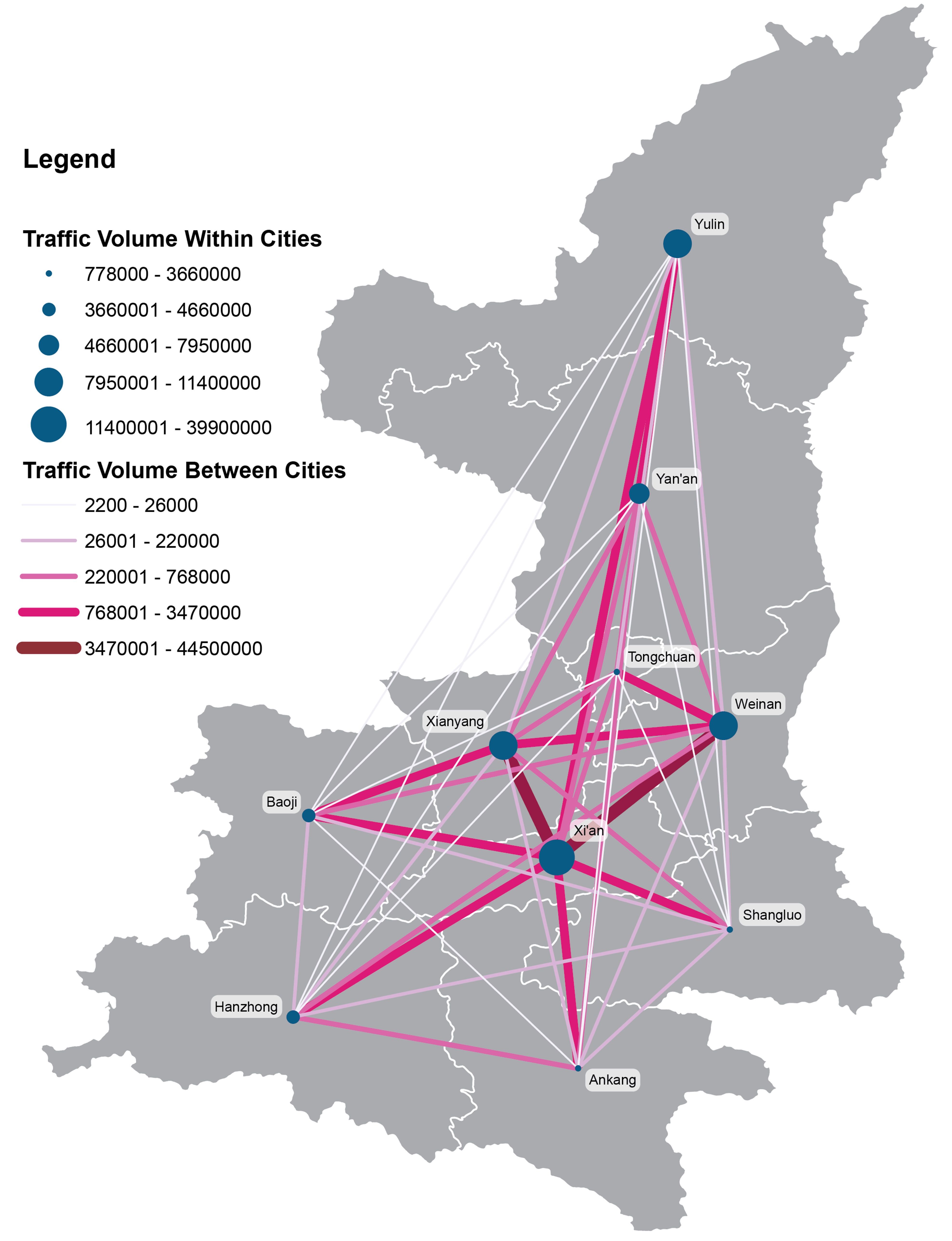}
		\caption{2014}
	\end{subfigure}
	\begin{subfigure}{0.45\linewidth}
		\centering
		\includegraphics[width=\linewidth]{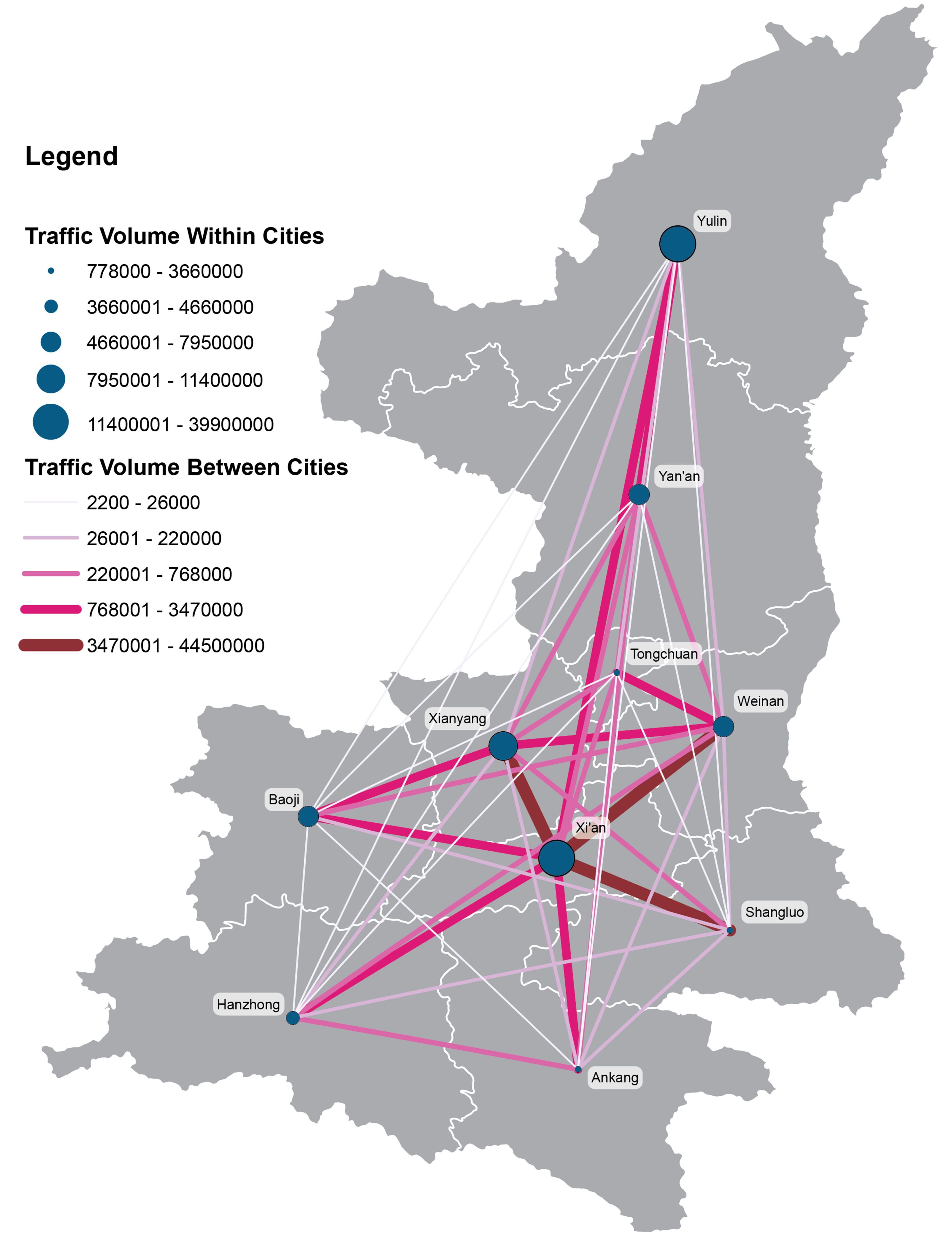}
		\caption{2015}
	\end{subfigure}
	\begin{subfigure}{0.45\linewidth}
		\includegraphics[width=\linewidth]{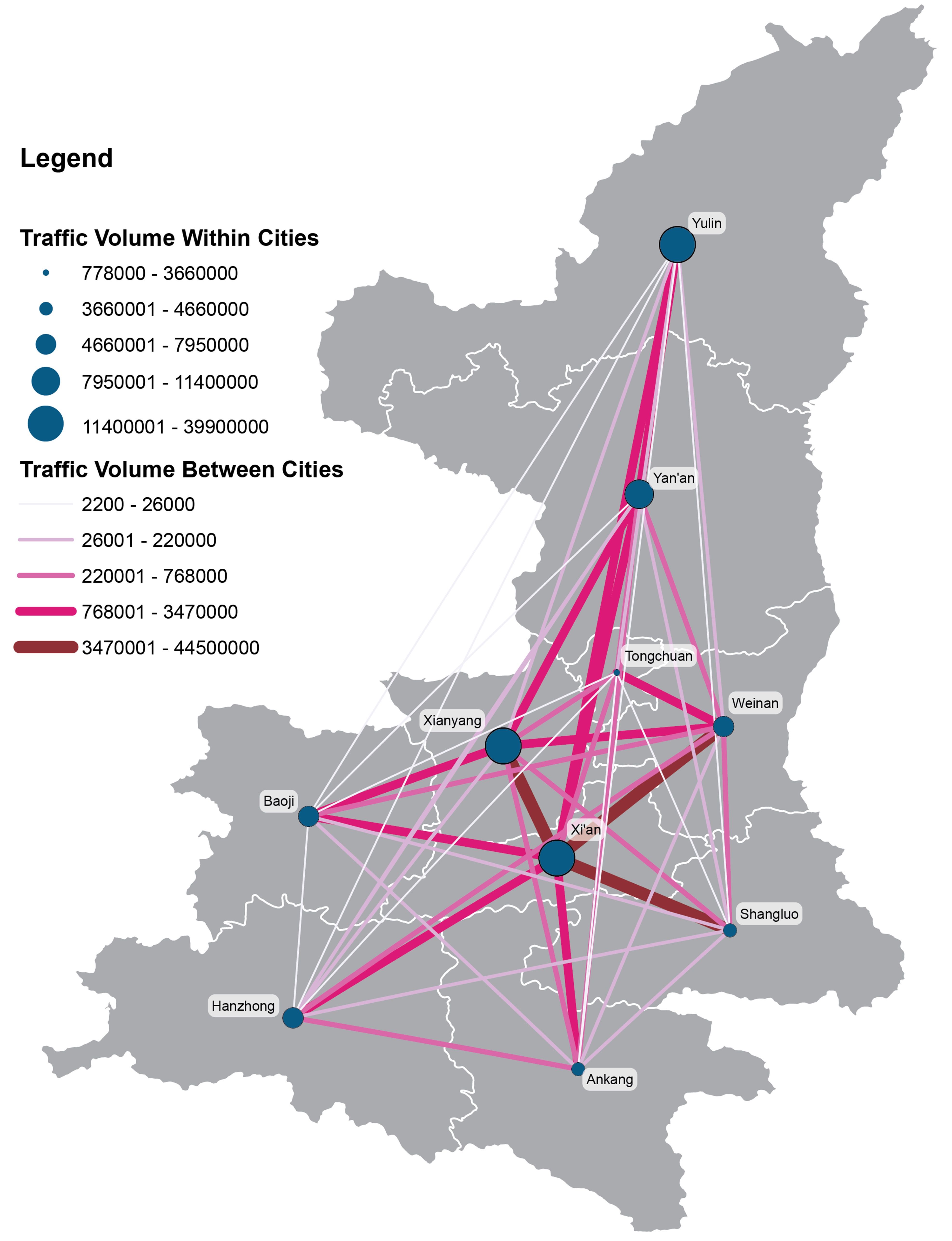}
		\caption{2016}
	\end{subfigure}
	\begin{subfigure}{0.45\linewidth}
		\includegraphics[width=\linewidth]{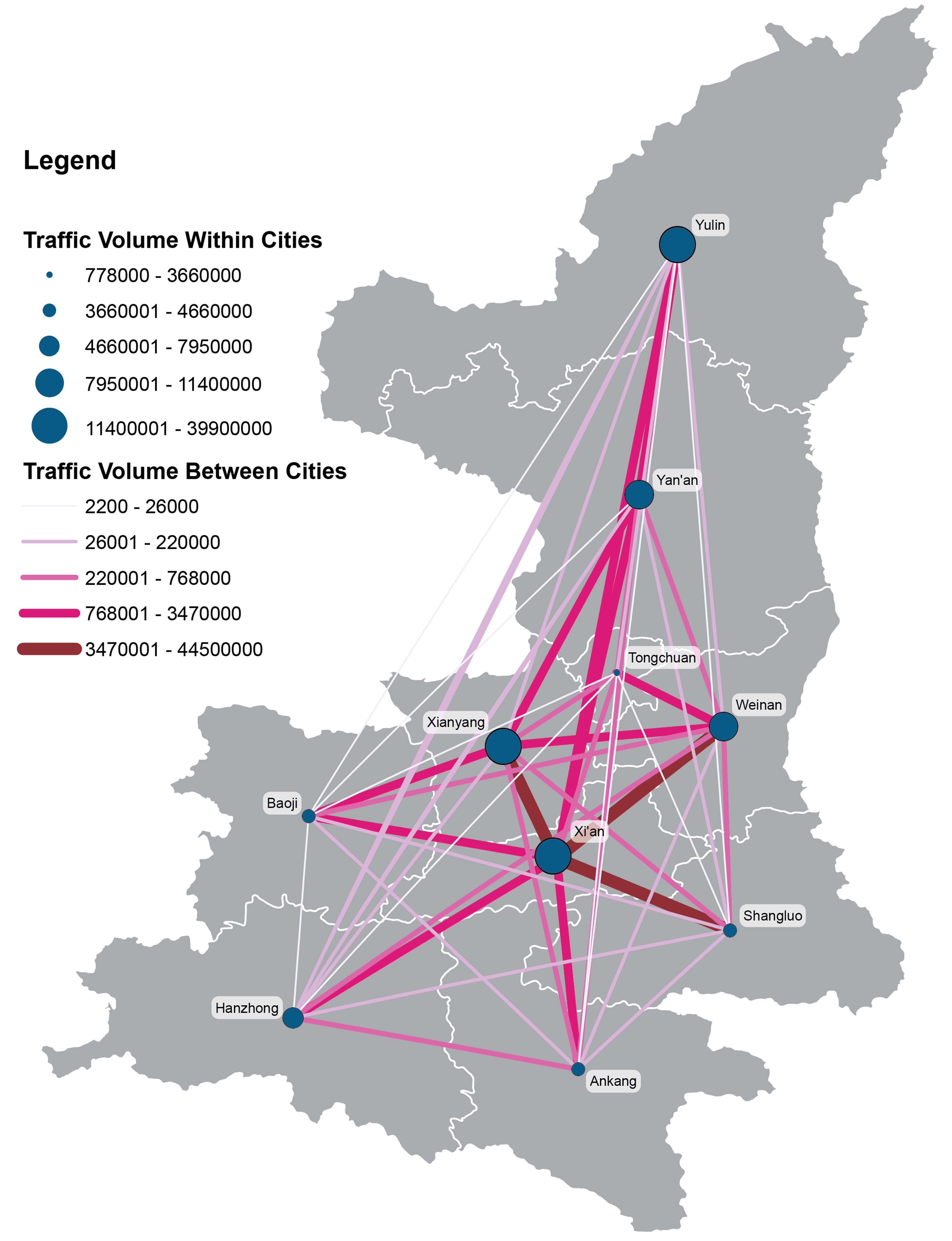}
		\caption{2017}
	\end{subfigure}
	\centering
	\caption{Mapping the annual traffic volumes of cars and buses among cities in Shaanxi province from 2014 to 2017. Note: The maps were generated using ArcMap version 10.6 and Adobe Illustrator CC version 20.}
	\label{fig:shaanxiflows}
\end{figure}

\begin{figure}[H]
	\centering
	\includegraphics[width=0.8\linewidth]{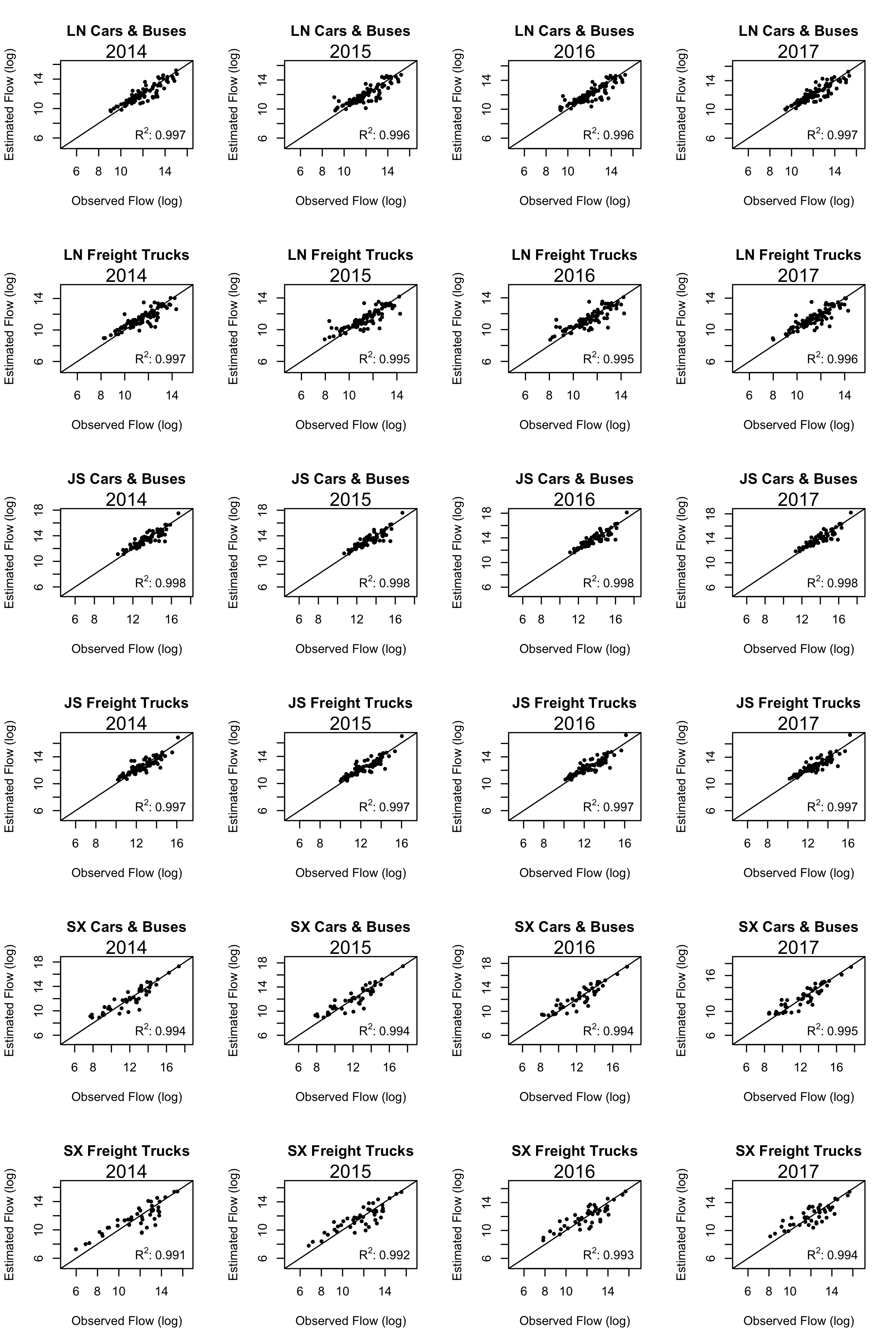}
	\caption{Relationships between the estimated and observed flow volumes of freight trucks, cars and buses in three provinces (LN: Liaoning; JS: Jiangsu; SX: Shaanxi) over four years using the gravity model. }
	\label{fig:city_fitG_bus_truck}
\end{figure}

\begin{figure}[h!]
	\begin{subfigure}[b]{0.4\linewidth}
		\includegraphics[width=\linewidth]{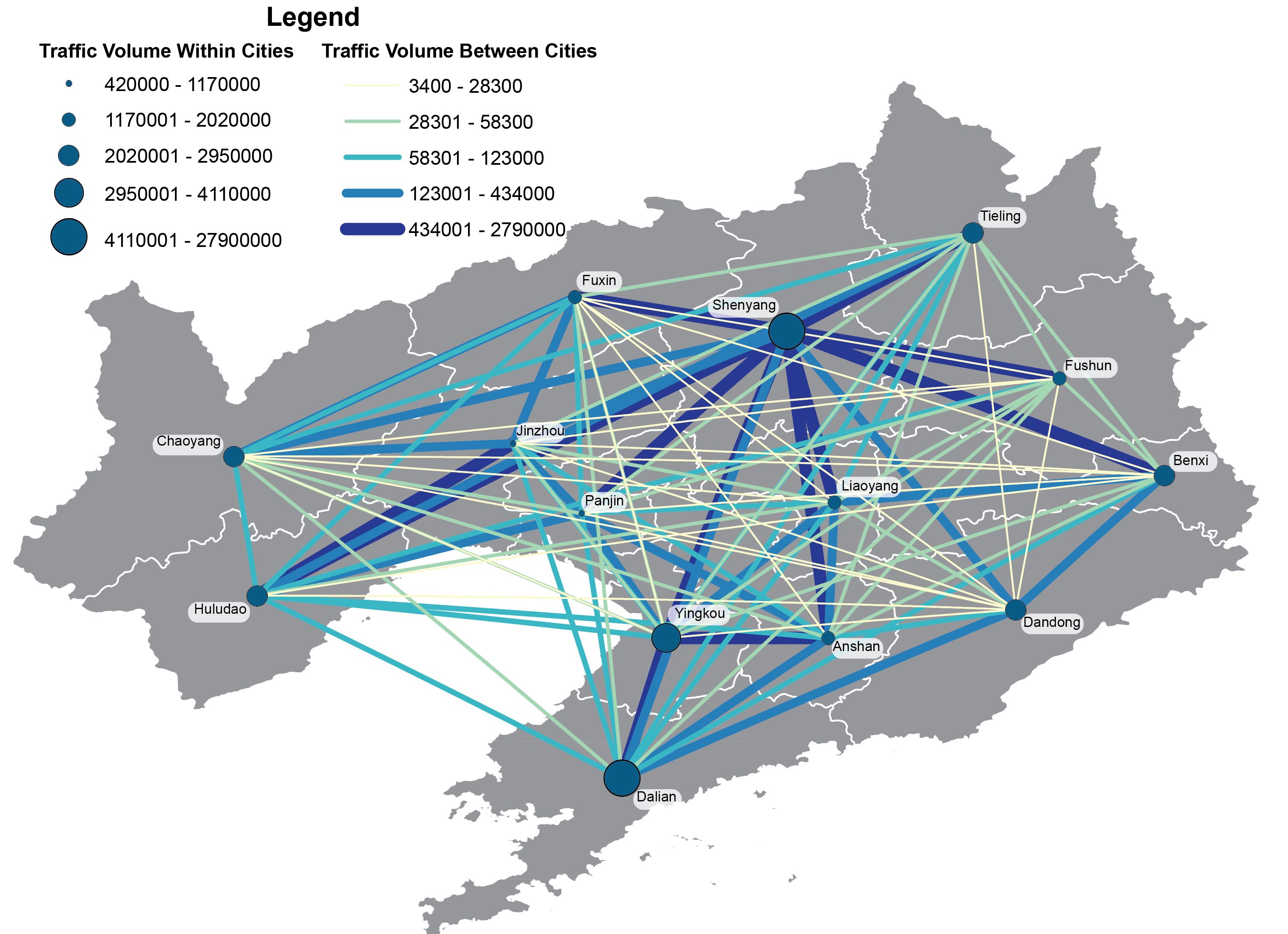}
		\caption{2014}
	\end{subfigure}
	\begin{subfigure}[b]{0.4\linewidth}
		\includegraphics[width=\linewidth]{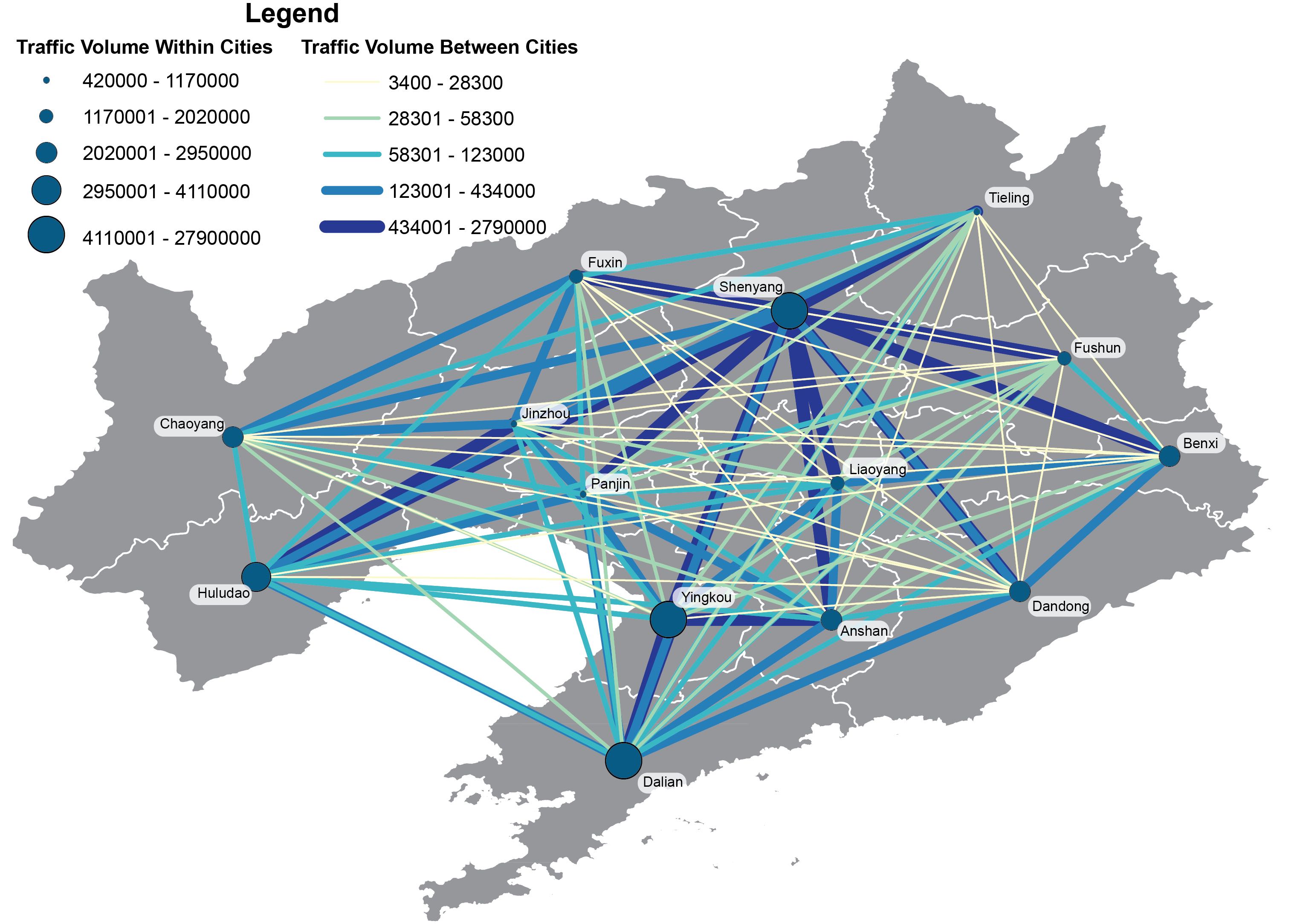}
		\caption{2015}
	\end{subfigure}
	\begin{subfigure}[b]{0.4\linewidth}
		\includegraphics[width=\linewidth]{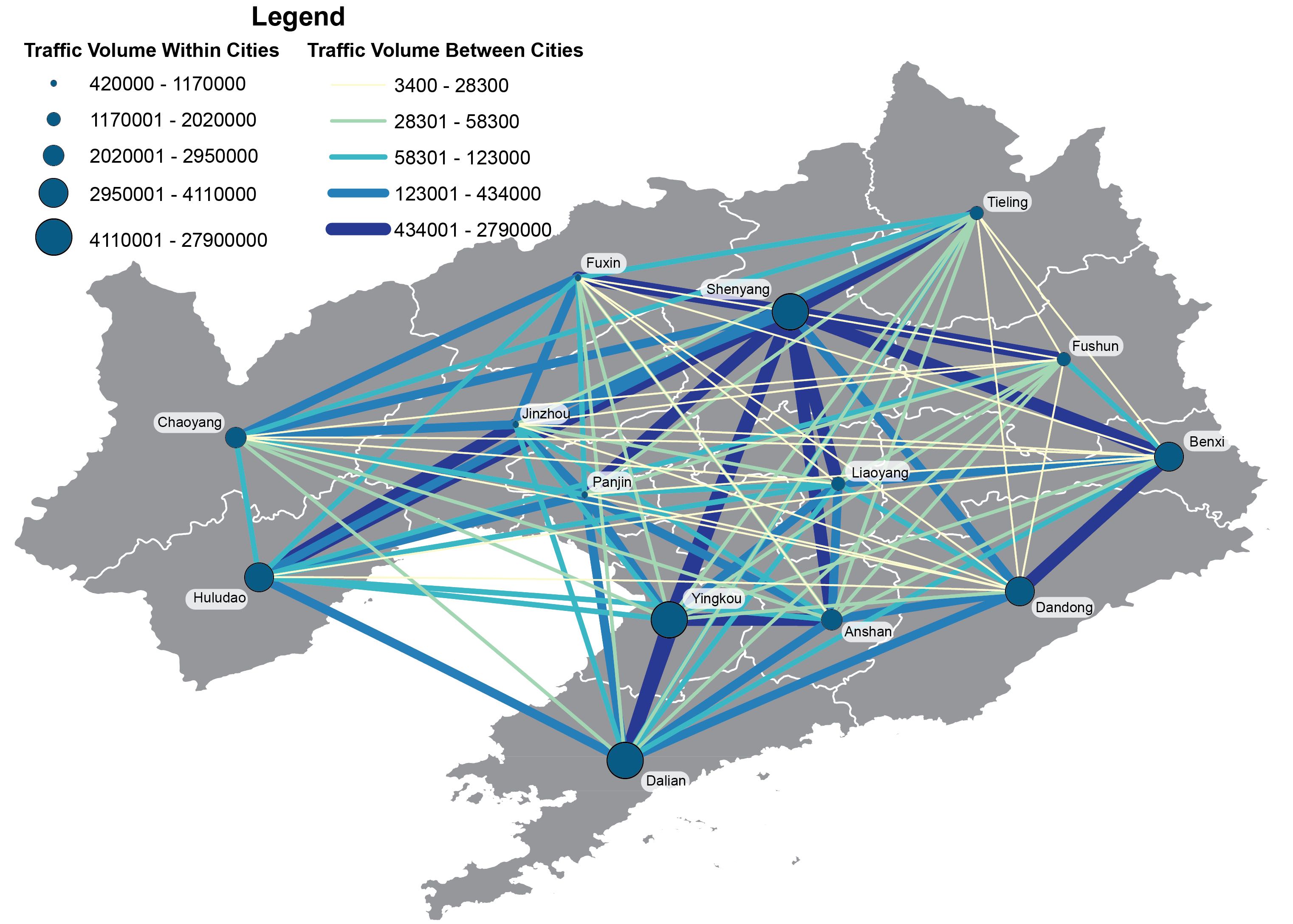}
		\caption{2016}
	\end{subfigure}
	\begin{subfigure}[b]{0.4\linewidth}
		\includegraphics[width=\linewidth]{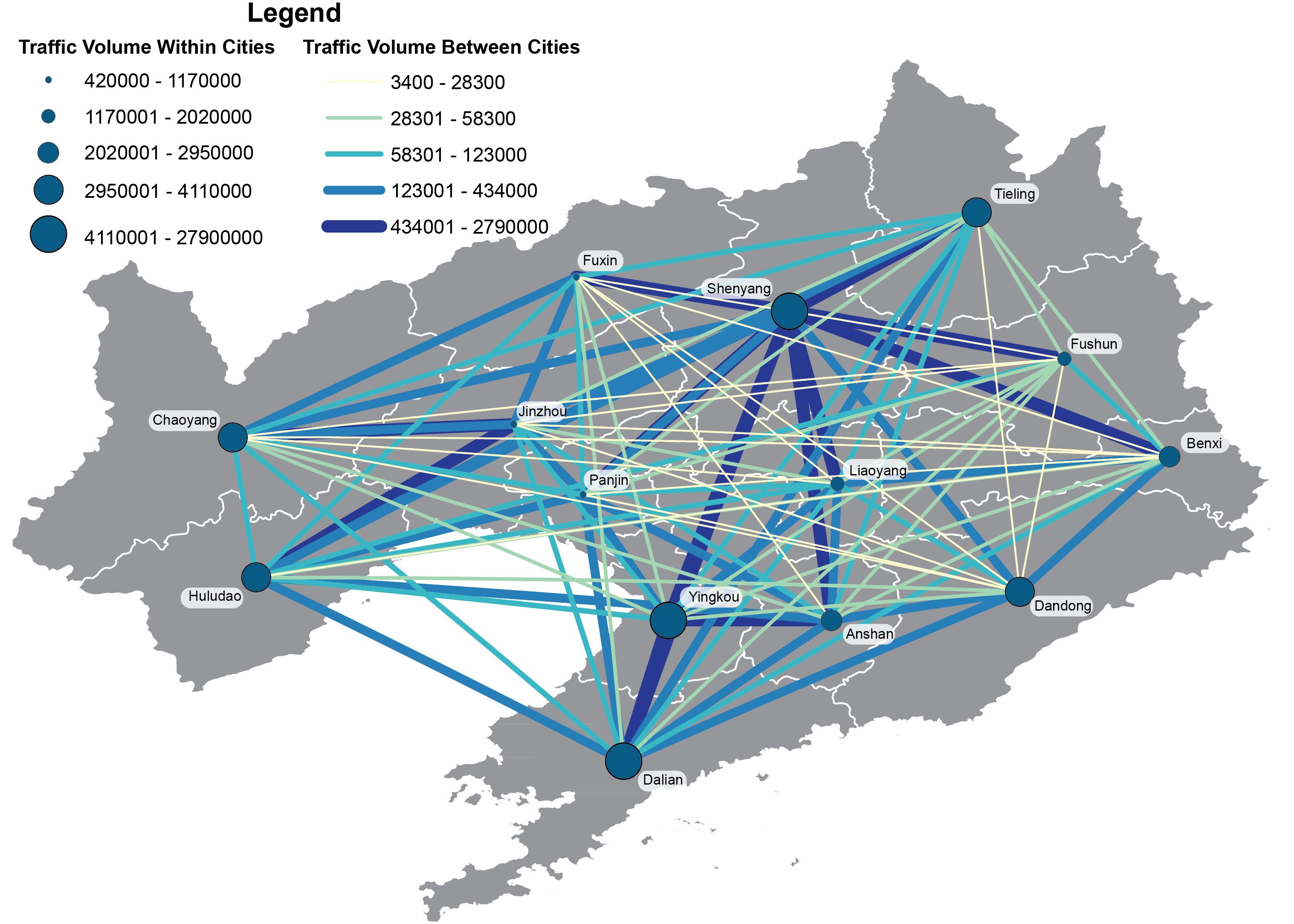}
		\caption{2017}
	\end{subfigure}
	\centering
	\caption{Traffic Volume of Cars and Buses in Liaoning. Note: The maps were generated using ArcMap version 10.6 and Adobe Illustrator CC version 20.}
	\label{fig:LNflows}
\end{figure}

\begin{figure}[h!]
	\begin{subfigure}[b]{0.4\linewidth}
		\includegraphics[width=\linewidth]{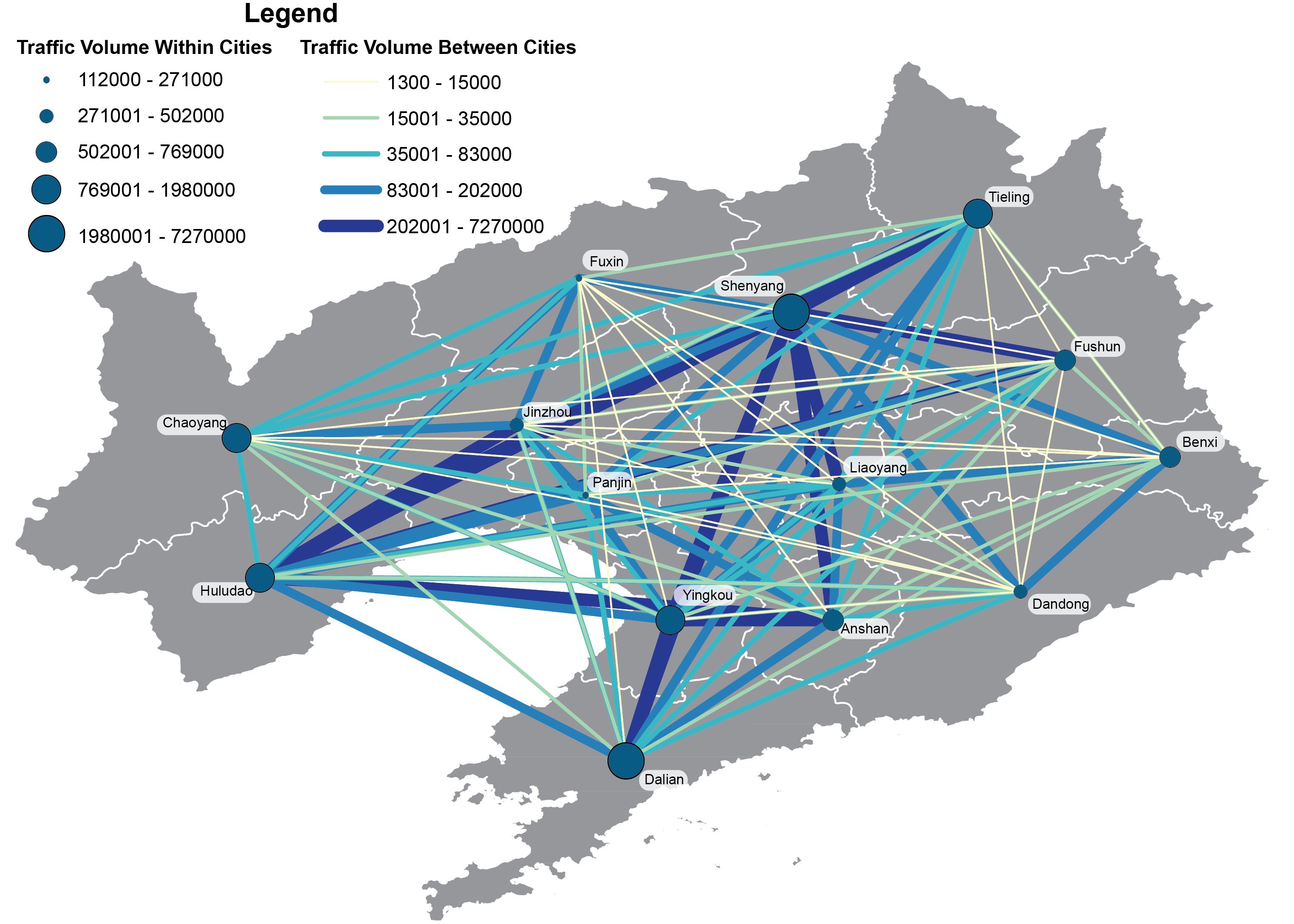}
		\caption{2014}
	\end{subfigure}
	\begin{subfigure}[b]{0.4\linewidth}
		\includegraphics[width=\linewidth]{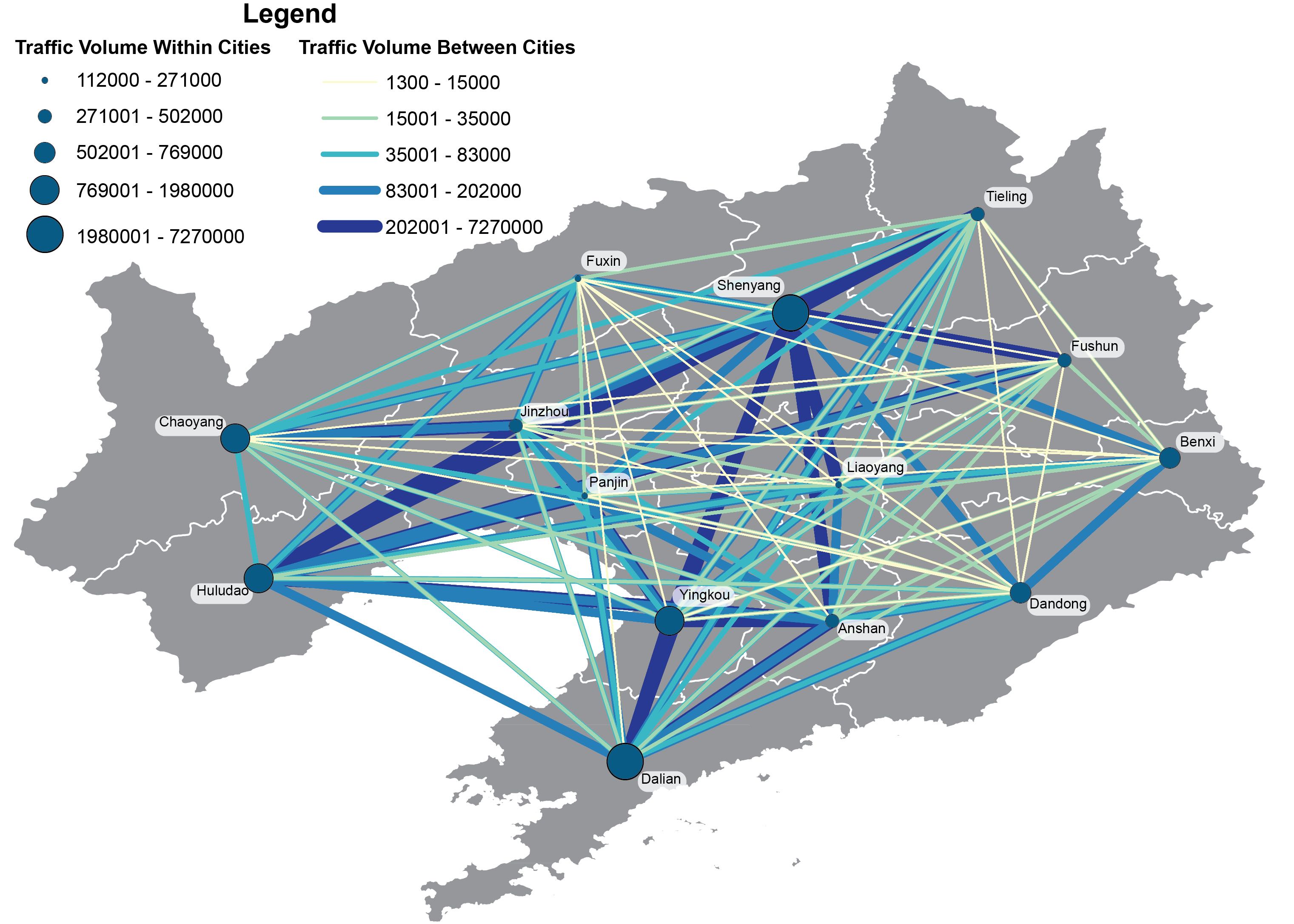}
		\caption{2015}
	\end{subfigure}
	\begin{subfigure}[b]{0.4\linewidth}
		\includegraphics[width=\linewidth]{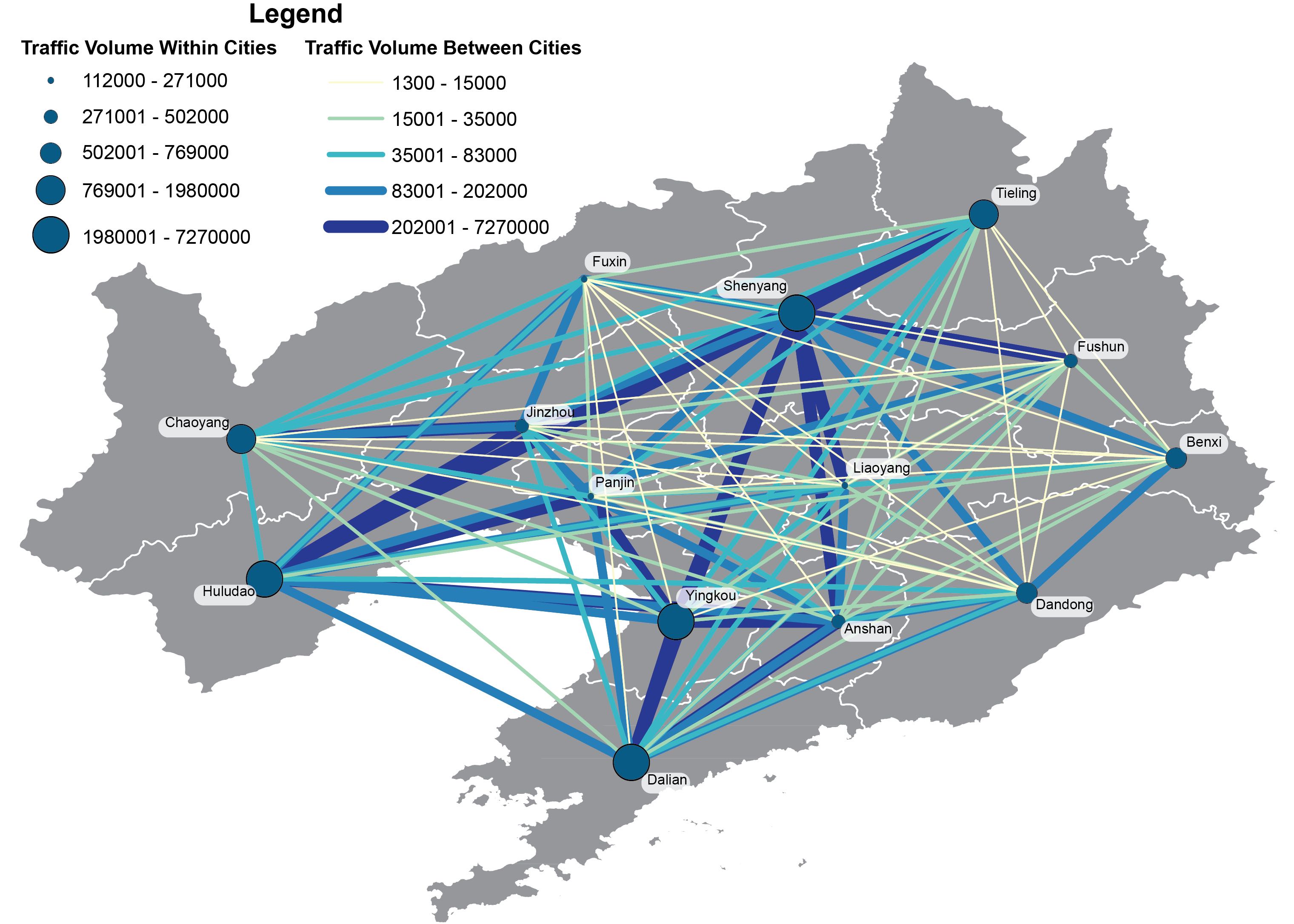}
		\caption{2016}
	\end{subfigure}
	\begin{subfigure}[b]{0.4\linewidth}
		\includegraphics[width=\linewidth]{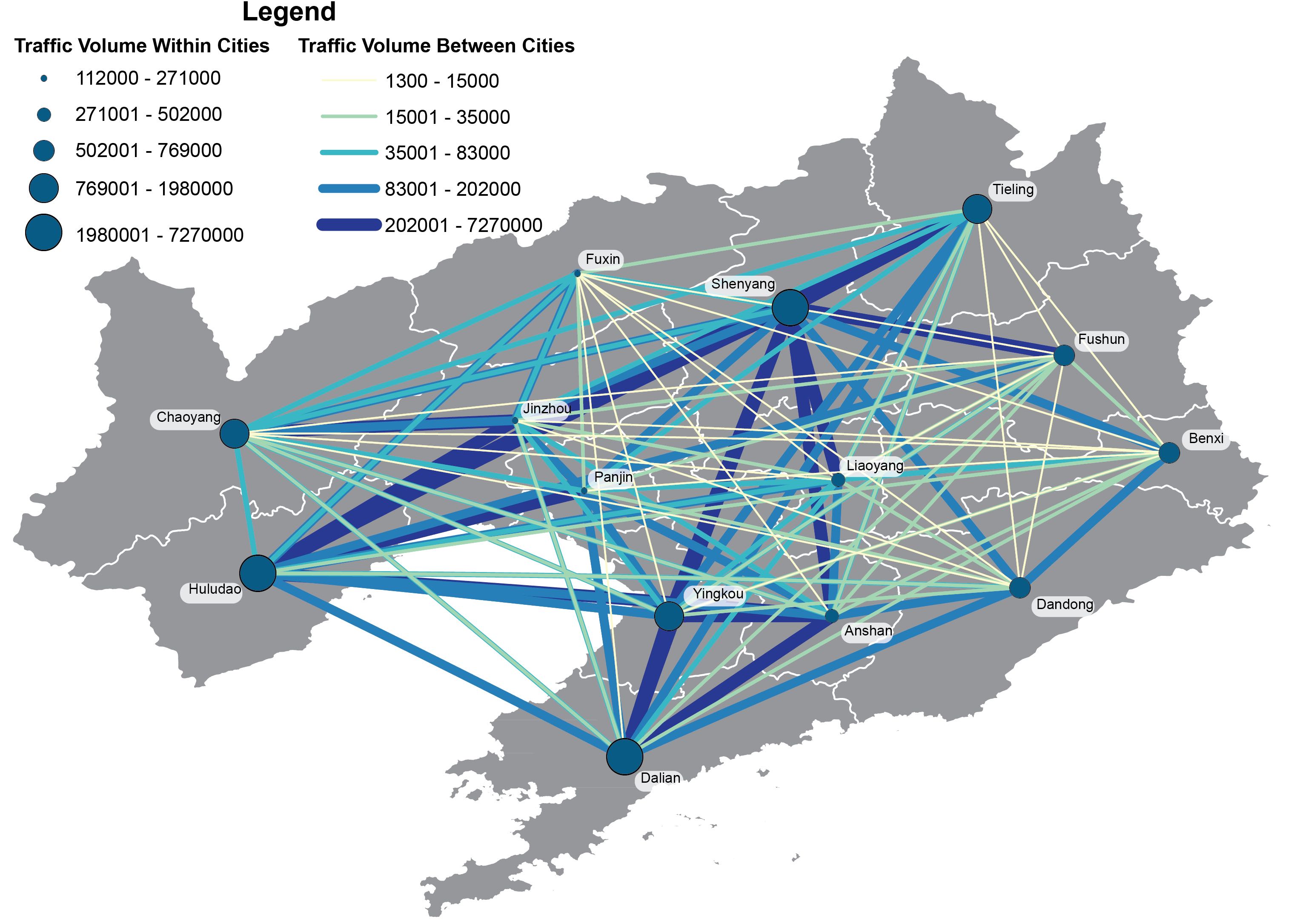}
		\caption{2017}
	\end{subfigure}
	\centering
	\caption{Traffic Volume of Freight Trucks in Liaoning. Note: The maps were generated using ArcMap version 10.6 and Adobe Illustrator CC version 20.}
	\label{fig:LNflows_truck}
\end{figure}

Next, we utilize the gravity model to estimate the city nodal attractiveness and the distance-decay effect on spatial interactions of cities based on the traffic-flow-weighted networks derived from cars $\&$ buses and freight trucks' entry and exit records along highways. Figure \ref{fig:shaanxiflows} shows the spatial distributions of the traffic volumes of cars and buses between cities in Shaanxi province. The magnitude of the traffic volumes of cars and buses within cities are larger than their pairwise inter-city traffic volumes for each category of vehicles (i.e., cars $\&$ buses, and trucks). The maps visualizing the traffic volumes of freight trucks in Shaanxi over four years can be found in Fig. \ref{fig:SXflows_truck}.  Similar maps of the other two provinces can be found in Supplement Figs.  \ref{fig:JSflows}, \ref{fig:JSflows_truck}, \ref{fig:LNflows}, and \ref{fig:LNflows_truck}.  The spatiotemporal distributions of traffic volumes in each province could reflect the highway construction and transportation trends in each province \cite{xiao2015trend,xiao2018index}. Figure \ref{fig:city_fitG_bus_truck} displays the relationships between the estimated and observed flow volumes of freight trucks, cars $\&$ buses in three provinces (LN: Liaoning; JS: Jiangsu; SX: Shaanxi) over four consecutive years using the gravity models. The goodness of fit R$^2$ of those models are all over 0.99 using the linear regression approach. The results indicate that the gravity model fits well in the regional land transportation patterns for cars $\&$ buses and freight trucks in all three provinces over the study period. 

\begin{figure}[H]
	\centering
	\includegraphics[width=0.95\linewidth]{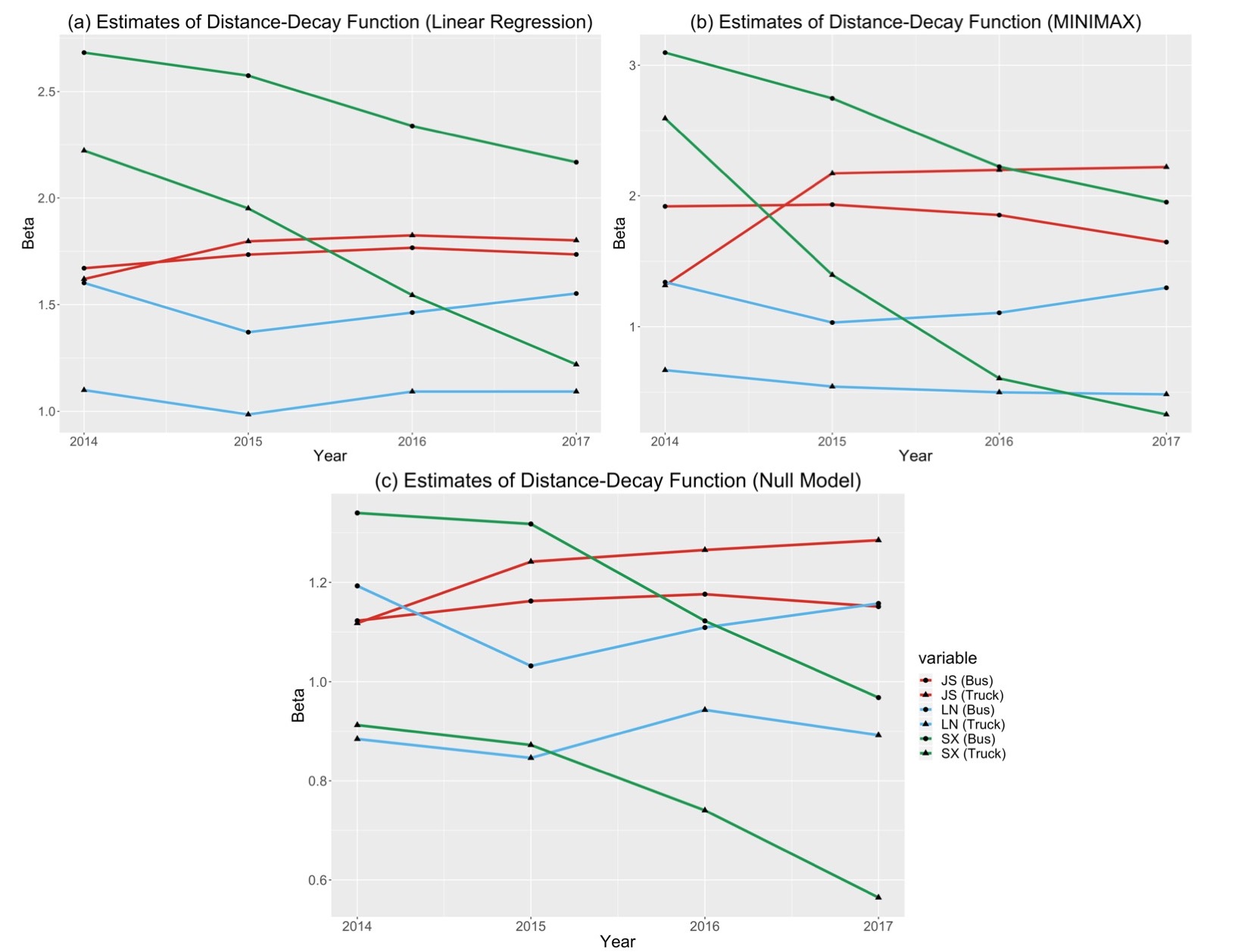}
	\caption{The temporal changes of distance decay effects on traffic-flow-based spatial interactions among cities in three provinces using three different parameter estimation approaches to the decay coefficient $\beta$. (a) using linear regression approach; (b) using MINIMAX linear programming approach; (c) using the null model.}
	\label{fig:distancedecay}
\end{figure}

In addition, we estimated the distance-decay coefficient $\beta$ in the gravity models for both categories of transportation networks using three different approaches: linear regression, linear programming (MINIMAX), and the Null model. Figure \ref{fig:distancedecay} shows the temporal changes of the distance-decay coefficient $\beta$ in three provinces using different parameter estimation approaches. Although we got different estimated absolute values of $\beta$ using varying parameter estimation strategies, the relative overall trend remains. Specifically, as the regional economy grows, we would expect more spatial interactions of people and goods (and other types of flows \cite{zhen2019analyzing}) among cities across different ranges of distances. Thus, it results in a smaller $\beta$ and vice versa. Furthermore, the rapid development of transportation infrastructures boosts the long-distance travels in a province. We found that the distance-decay effects decreased significantly in Shaanxi province and aligned well with its fast economic growth trend in the study period (see Figure \ref{fig:GDP_transportation}). In contrast, the regional economy in Northeast China grew slowly in recent years and even decreased in several cities. Liaoning province was selected as a representative province to illustrate this trend. From year 2015 to 2016, most cities' GDP decreased which may limit some long-distance travels of passengers and goods. Thus, the distance-decay coefficient $\beta$ increased from 1.03 to 1.11 for traffic flows of cars and buses using the null model estimation approach, and increased from 0.85 to 0.94 for flows of freight trucks respectively. Besides, the distance decay effects of traffic flows in Jiangsu province didn't change much while it kept a stable economic growth across all cities over years. The spatial concentration of economic activities and interactions within the southern part of Jiangsu but fewer transportation interactions between the southern and the northern parts may contribute to the high distance-decay effect (as shown in Figure \ref{fig:JSflows}). Moreover, we found that the distance-decay coefficients $\beta$ for the truck flow networks are smaller than those of the cars $\&$ buses flow networks in Shaanxi and Liaoning provinces, where long-distance travels of freight trucks are important components of regional economic activities.

\subsection*{Network measures and sub-network structure }
\begin{figure}[H]
	\centering
	\includegraphics[width=\linewidth]{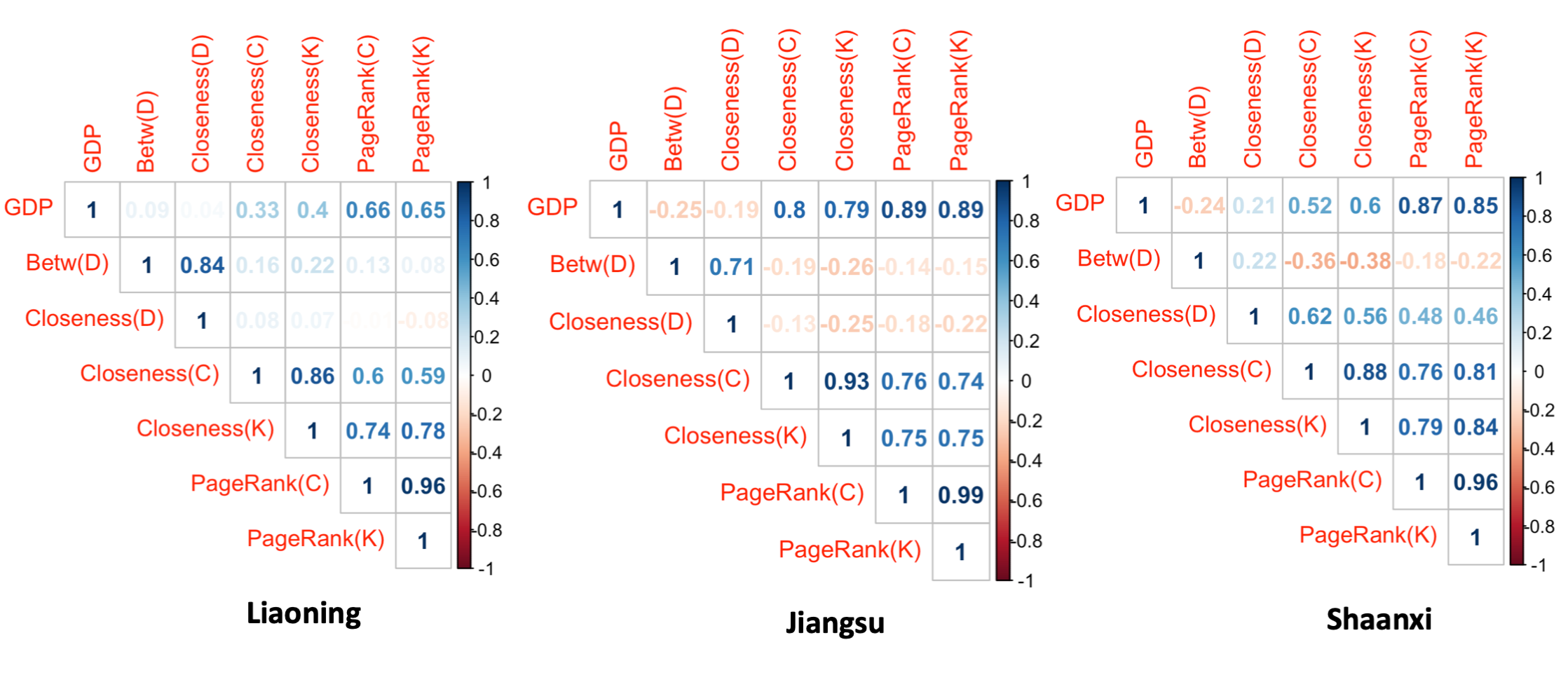}
	\caption{The Pearson's correlation coefficients between city GDP value, betweenness, closeness centrality measures and the PageRank index in transport flow networks of cars $\&$ buses (C) and trucks (K) in three provinces. The Betw (D) and Closeness (D) measures are calculated using the spatial interaction networks of cities with the inter-city distances as edge weights.}
	\label{fig:centrality_cor}
\end{figure}

Network science methods are useful in uncovering inherent characteristics of transportation networks and the spatial structure of cities and regions \cite{chi2016uncovering,zhen2019analyzing}. A variety of network indicators such as Centrality, PageRank, LeaderRank and many others have been proposed for identifying influential nodes in complex networks \cite{freeman1978centrality,chen2012identifying,lu2016vital}.  We constructed three spatial interaction networks of cities based on different weight choices: (1) the physical distances between city centers $G_D<V, E, W_D>$; (2) the inter-city flow volumes of cars and buses $G_C<V, E, W_C>$; and (3) the inter-city flow volumes of freight trucks $G_K<V, E, W_K>$. Then, the betweenness and closeness centrality measures were computed in the network $G_D$, which help understand the spatial structure of cities in the physical space. The betweenness quantifies node importance based on how often shortest paths pass through a specific node, but it needs to incorporate human population distribution and the distance decay function to better explain traffic flow distribution \cite{gao2013understanding}. As for the flow-weighted networks $G_C$ and $G_K$, the closeness centrality and the weighted PageRank \cite{brin1998anatomy,page1999pagerank,mao2015quantifying} were computed to measure the node importance. Then, we computed the Pearson's correlation coefficient between city GDP and each of these node importance measures. As shown in Figure \ref{fig:centrality_cor}, the city PageRank measures in the networks of cars $\&$ buses $PageRank(C)$ and trucks $PageRank(K)$ strongly positively correlate with city GDP in all three provinces. The flow-weighted closeness centrality $Closeness(C)$ and $Closeness(K)$ are positively correlated with city GDP whereas neither $Closeness(D)$ nor $Betw(D)$ centrality in physical space had a significant correlation with city GDP. In sum, the weighted network measures using the traffic flows correlate better with regional economy than that using the physical distance-based ones. 

\begin{figure}[H]
	\begin{subfigure}{0.48\linewidth}
		\centering
		\includegraphics[width=\linewidth]{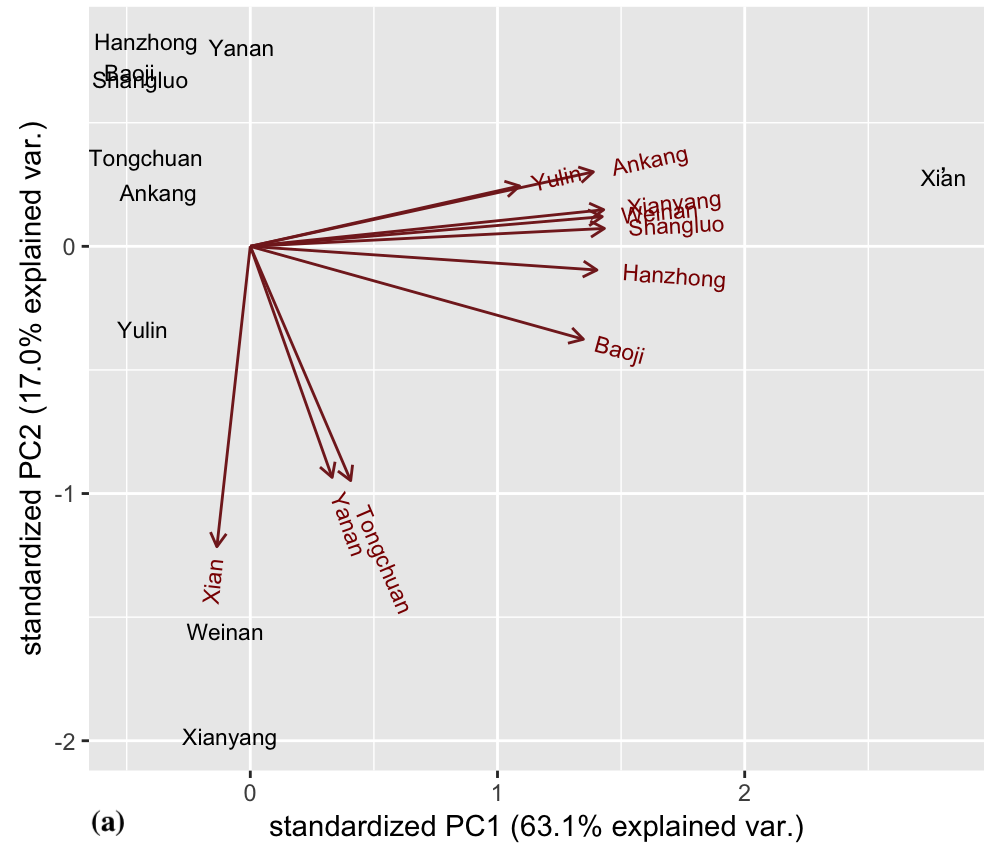}
	\end{subfigure}
	\begin{subfigure}{0.48\linewidth}
		\includegraphics[width=\linewidth]{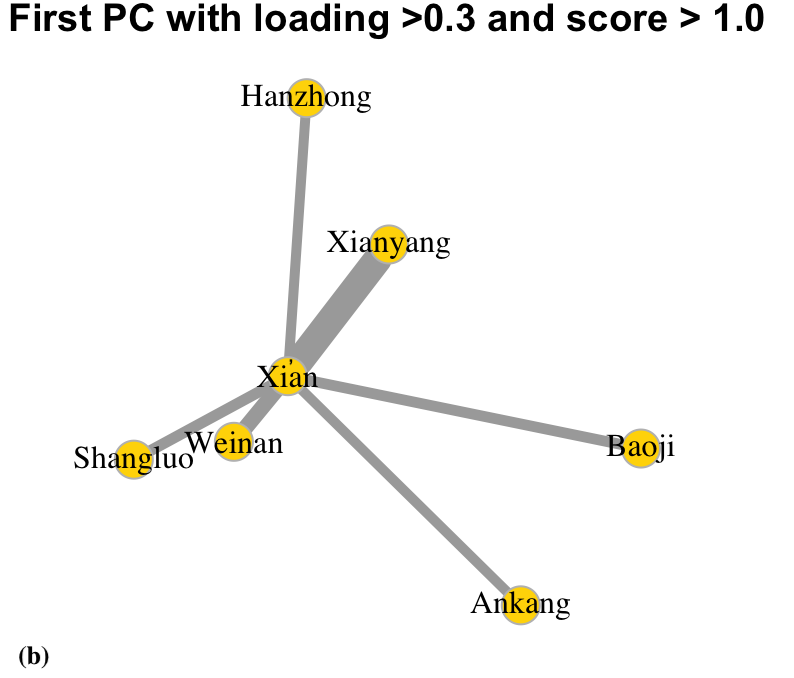}
	\end{subfigure}
	\centering
	\caption{The PCA analysis results of spatial interaction networks of buses and cars in Shaanxi province. (a) the city coordinates using the first two principle components PC1 and PC2; the axes in red represent original coordinate space. (b) the graph structure of first principle component PC1 with factor loading > 0.3 and component score > 1.0. Note: The figures were generated using RStudio version 1.2.}
	\label{fig:pca_SX}
\end{figure}

\begin{figure}[htb!]
	\begin{subfigure}[b]{0.45\linewidth}
		\includegraphics[width=\linewidth]{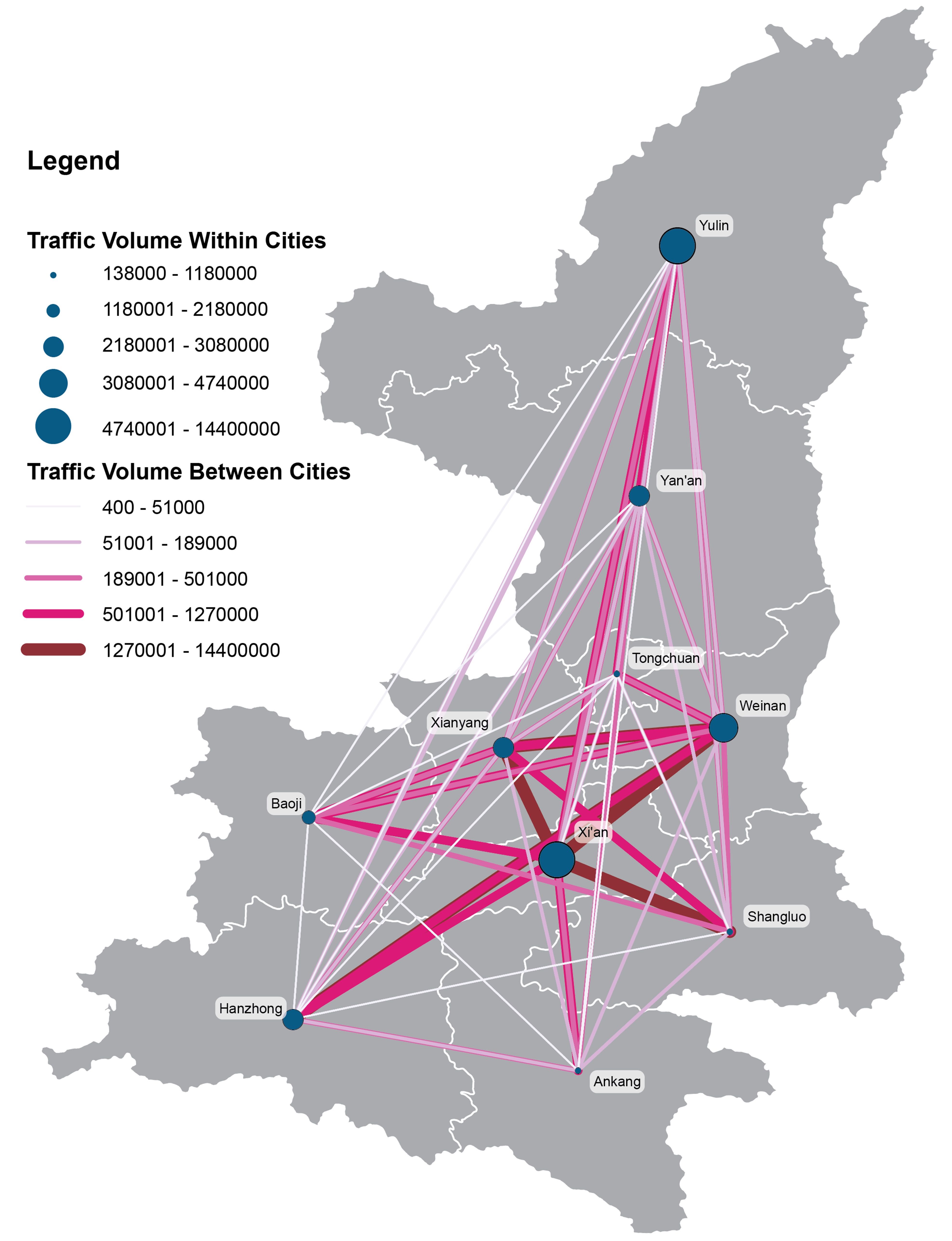}
		\caption{2014}
	\end{subfigure}
	\begin{subfigure}[b]{0.45\linewidth}
		\includegraphics[width=\linewidth]{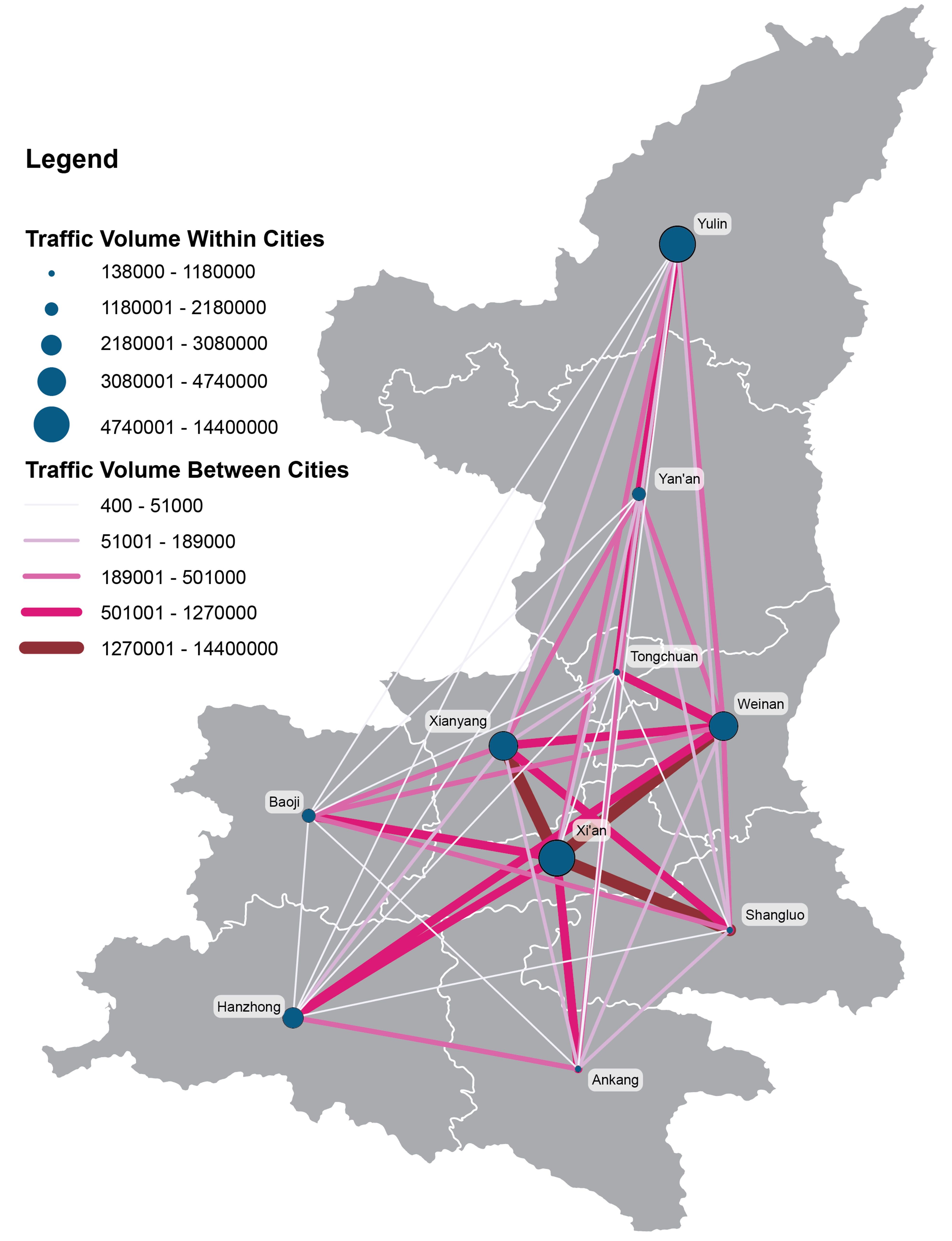}
		\caption{2015}
	\end{subfigure}
	\begin{subfigure}[b]{0.45\linewidth}		
		\includegraphics[width=\linewidth]{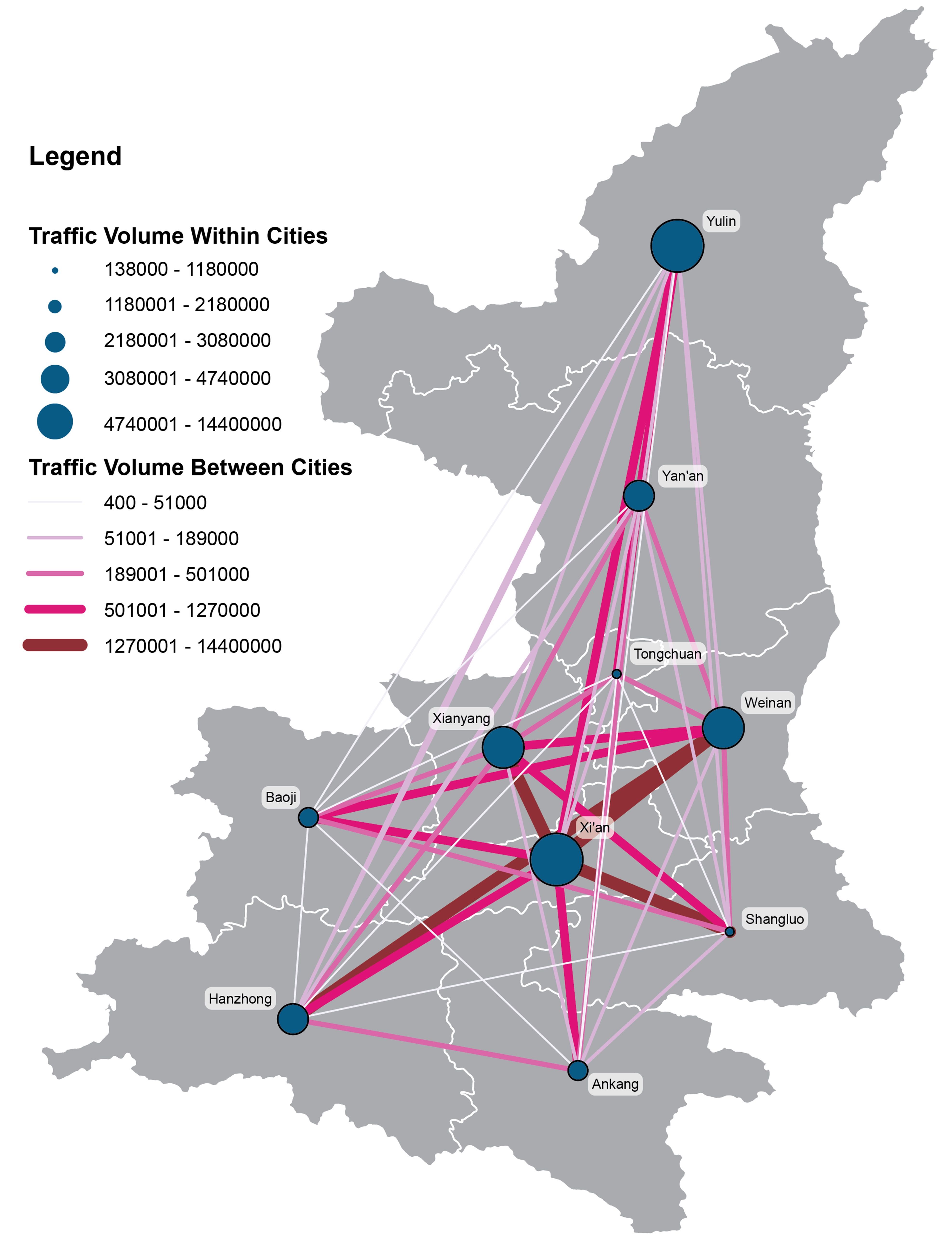}
		\caption{2016}
	\end{subfigure}
	\begin{subfigure}[b]{0.45\linewidth}
		\includegraphics[width=\linewidth]{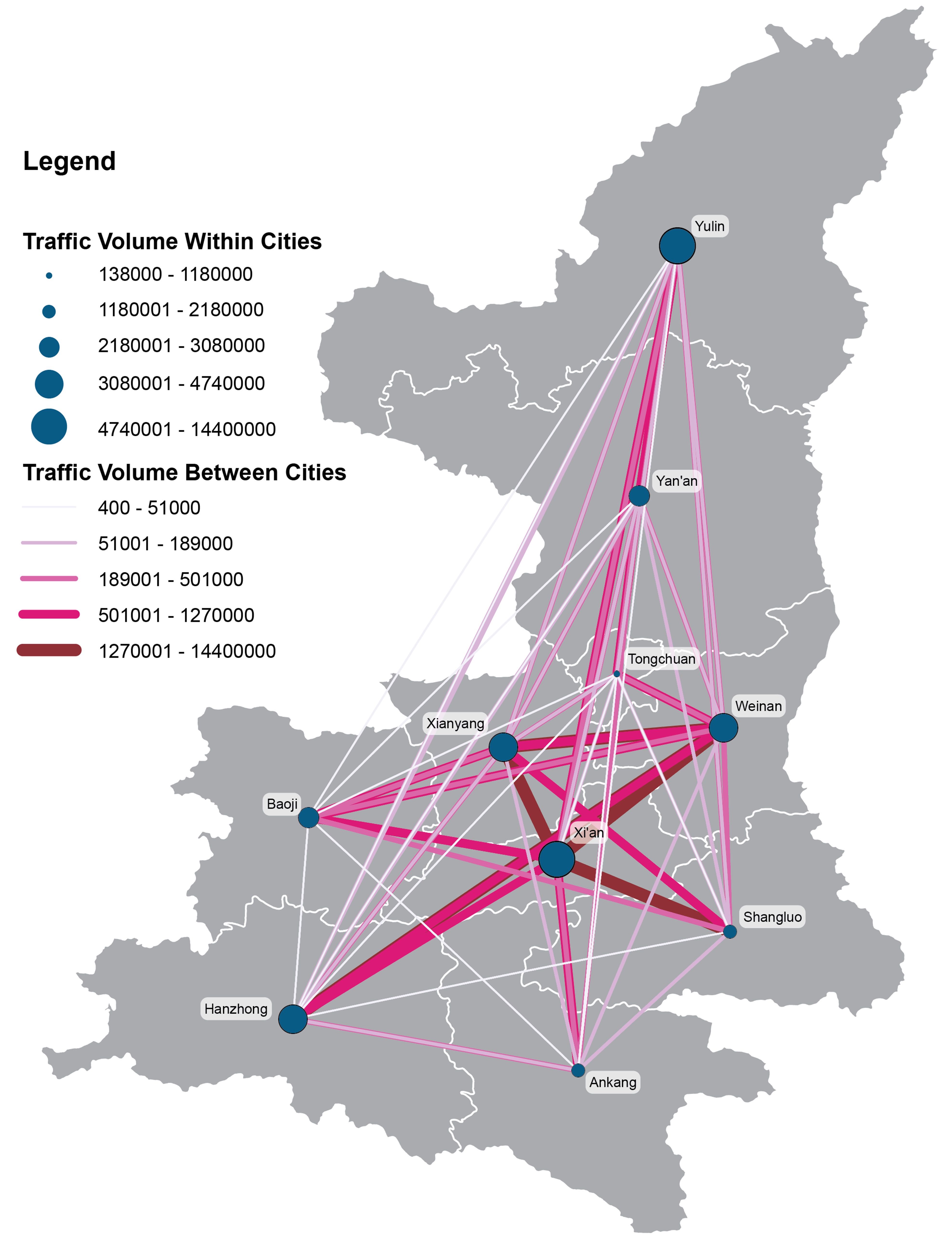}
		\caption{2017}
	\end{subfigure}
	\centering
	\caption{Traffic Volume of Freight Trucks in Shaanxi. Note: The maps were generated using ArcMap version 10.6 and Adobe Illustrator CC version 20.}
	\label{fig:SXflows_truck}
\end{figure}

\begin{figure}[h!]
	\begin{subfigure}{0.48\linewidth}
		\centering
		\includegraphics[width=\linewidth]{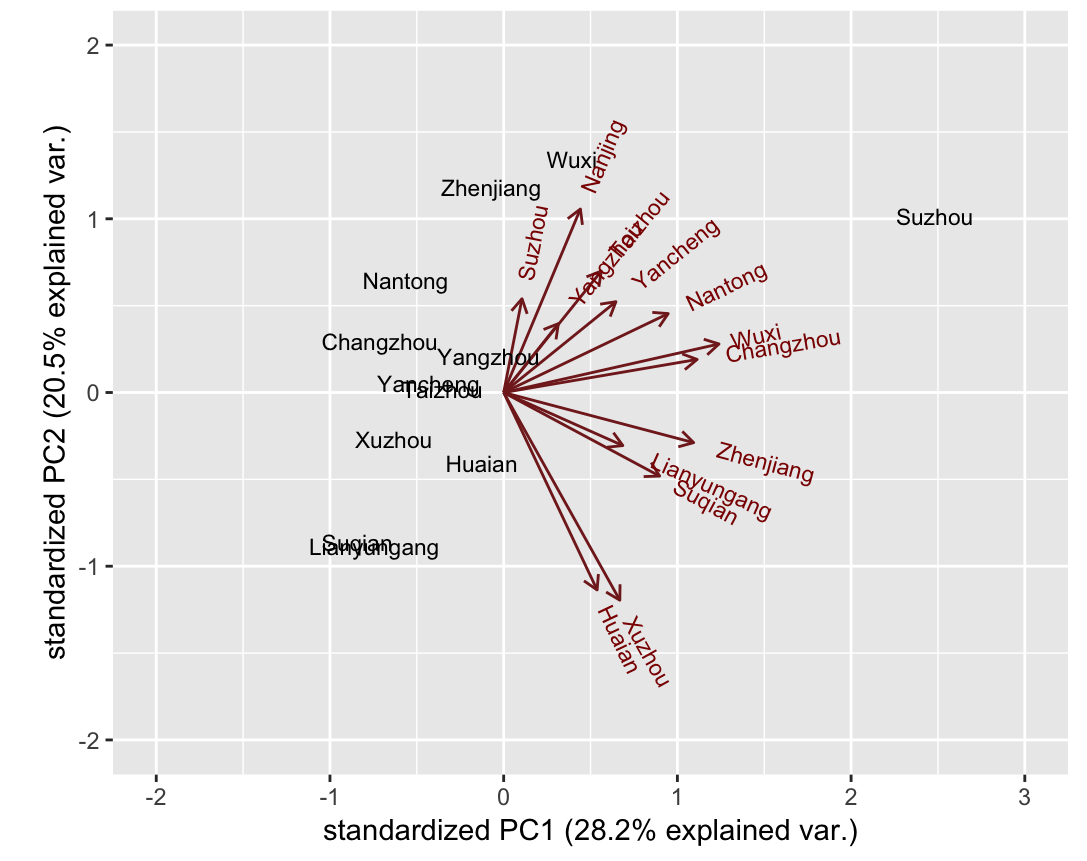}
	\end{subfigure}
	\begin{subfigure}{0.48\linewidth}
		\includegraphics[width=\linewidth]{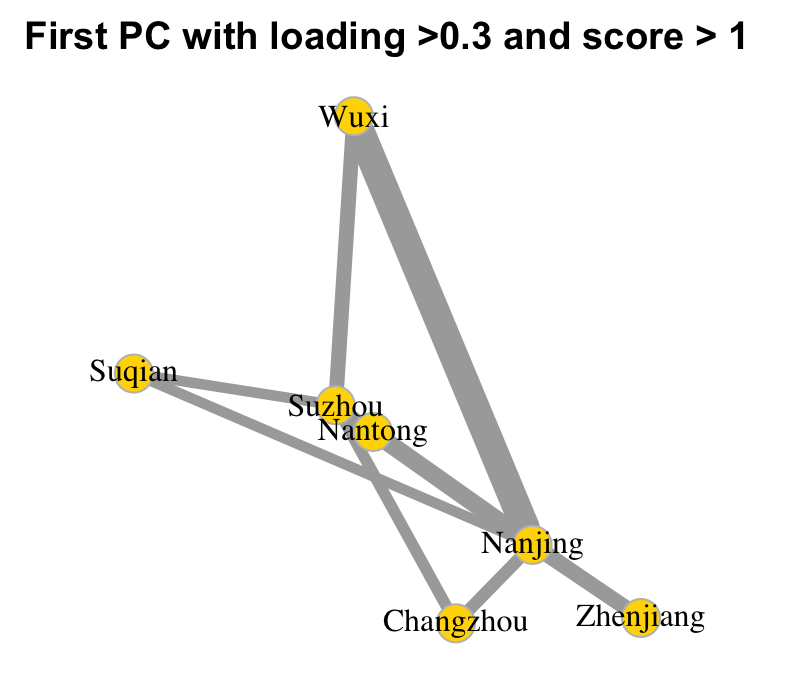}
	\end{subfigure}
	\centering
	\caption{The PCA analysis results of spatial interaction networks of buses and cars in Jiangsu. Note: The figures were generated using RStudio version 1.2.}
	\label{fig:pca_JS}
\end{figure}

\begin{figure}[h!]
	\begin{subfigure}{0.48\linewidth}
		\centering
		\includegraphics[width=\linewidth]{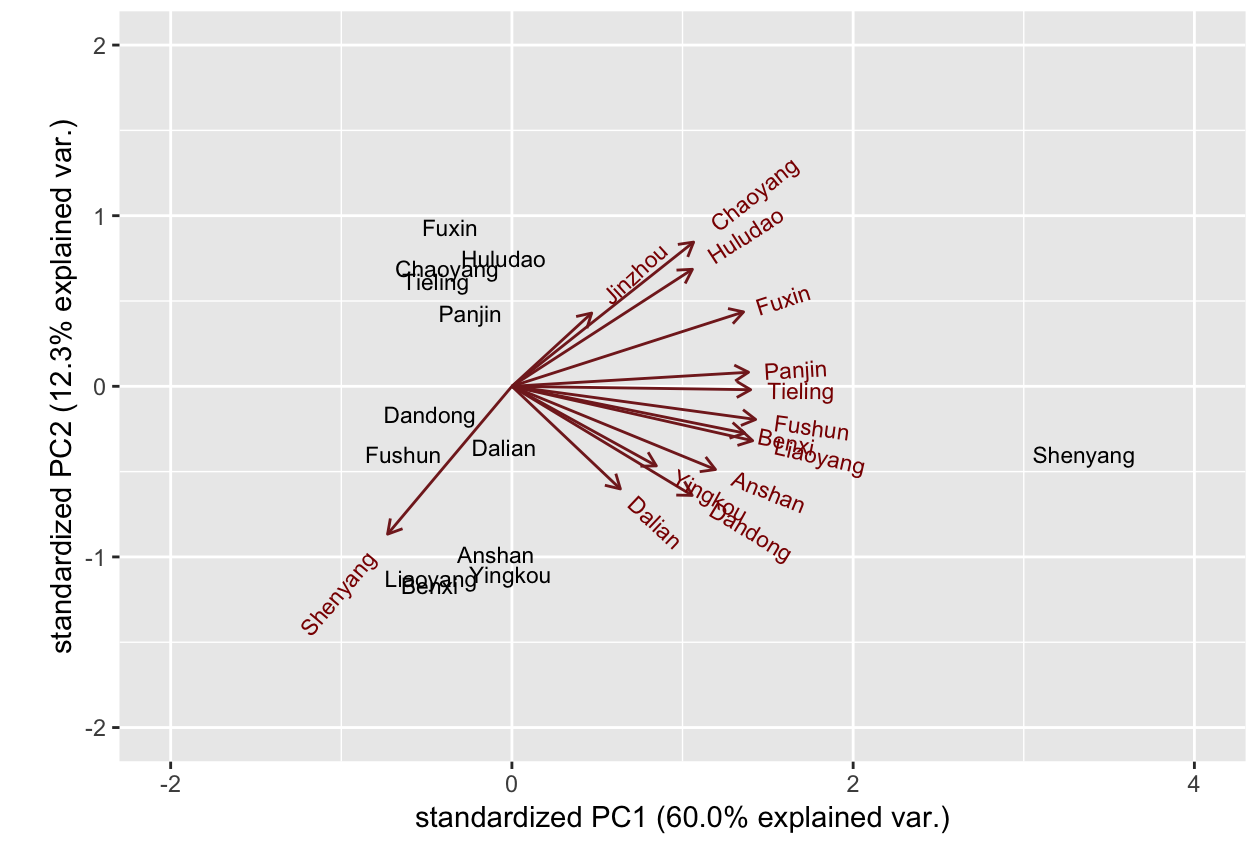}
	\end{subfigure}
	\begin{subfigure}{0.48\linewidth}
		\includegraphics[width=\linewidth]{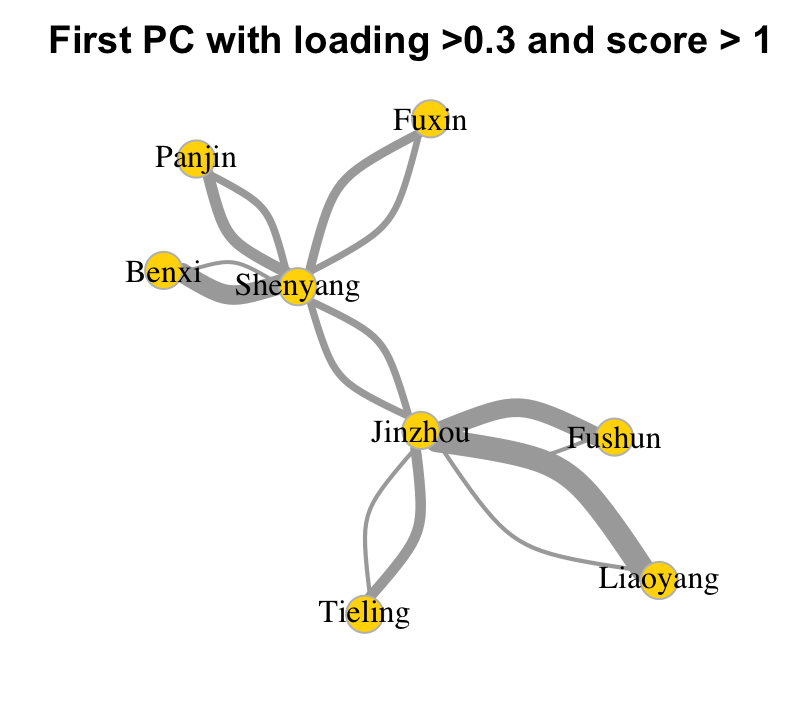}
	\end{subfigure}
	\centering
	\caption{The PCA analysis results of spatial interaction networks of buses and cars in Liaoning. Note: The figures were generated using RStudio version 1.2.}
	\label{fig:pca_LN}
\end{figure}

In addition, by applying principle component analysis (PCA) on the spatial interaction network matrices, we extract the subsystem of flows with a large portion of the total variance of interactions among cities. The first few components represent a substantial part of the total variance, which could reveal prominent regional transportation connection patterns. Taking the inter-city spatial interaction network of Shaanxi as an example, using the observed flow volumes of cars and buses in year 2014 as the weights, the PCA results in Figure \ref{fig:pca_SX}(a) show that first two principle components already account for over 80$\%$ (PC1: 63.1$\%$ and PC2: 17.0$\%$) of the total variance of the inter-city flows. Xi'an, as the provincial capital city of Shaanxi province, has the largest standardized component score for the first principle component. By linking each group of city destinations (filtered by the factor loadings) to a common set of strongly connected origins (filtered by component scores) using the Goddard's approach \cite{goddard1970functional}, a dominant nodal sub-system is extracted. As shown in Figure \ref{fig:pca_SX}(b) and Figure \ref{fig:shaanxiflows}, this sub-system links several important cities in the central and southern parts of Shaanxi province through the highway transportation system, including Xi'an, Xianyang, Weinan, Baoji, Shangluo, Ankang, and Hanzhong. Interestingly, the same dominant nodal sub-system is also extracted using the freight truck traffic volumes as the weights for the inter-city spatial interaction networks as shown in Figure \ref{fig:SXflows_truck}. This finding confirms the strong regional spatial connections of people and goods to support economic development in this area. Similarly, the PCA analysis results for extracting the dominant nodal sub-systems of Jiangsu province and Liaoning province are shown in Figure \ref{fig:pca_JS} and \ref{fig:pca_LN}. The sub-network structure reflects the complexity and the spatial connection characteristics of the regional economy.

\section*{Discussion and Conclusion} \label{section:discussion}

When estimating the city GDP from traffic flow variables, the existence of multicollinearity among independent variables may change the coefficient estimates of the MLR model erratically in response to small changes in the data. We used the variance inflation factors (VIF), which is a measure of how much the variance of the estimated regression coefficient $b_k$ (where $k$ is the index of a predictor) is "inflated" by the existence of correlation among the predictor variables in the model. 

\begin{equation} 
VIF_k = \frac{1}{1-R^2_k}
\end{equation}
Where $R^2_k$ is the R-squared obtained by regressing the $k_{th}$ predictor on the remaining predictors.

The VIF scores for the predictor variables incoming flow ($I_C$), outgoing flow ($O_C$) of cars and buses are very high (> 2000) across all three provinces. As the original multi-linear regression model includes the ratio of incoming/outgoing flow ($R_C$), such high multicollinearity is expected. After removing the two variables with largest VIF scores in each regression model for predicting the city GDP, the R-squared values slightly reduced: from 0.934 to 0. 909 (Liaoning), from 0.892 to to 0.889 (Jiangsu), and from 0.967 to 0.955 (Shaanxi), respectively.

Moreover, by calibrating the model penalty term of  L1-norm and L2-norm using empirical data in each province, we derived the regularized regression results using the Ridge and LASSO approaches. As shown in Table \ref{table:regularization}, both Ridge and LASSO methods got a high goodness of fit in the predictive modeling of city GDP: R-squared of 0.887 (Ridge) and 0.932 (LASSO) in Liaoning province, 0.888 (Ridge) and 0.882 (LASSO) in Jiangsu province, and  0.961 (Ridge) and 0.965 (LASSO) in Shaanxi province. The R-squared results are almost as good as the non-regularized MLR results using all features, but the regularized regression models are more stable in the cross-validation experiments although with some additional cost of bias. Regarding the feature/predictor selection, each province has its own unique combination of transport flow predictors to its city economy. The models also identify the prominent predictor (i.e.,  the intracity flow of cars and buses $N_C$) that has a large positive impact on their GDP values but with different magnitudes of coefficients across three provinces. The detailed model coefficient results can be found in Table \ref{table:regularization}. Interestingly, if we combine all the cities' data from three provinces into one linear model, the top two selected predictors are the intercity incoming flow of trucks ($I_{K}$) and the intracity flow of trucks ($N_{K}$) for both Ridge and LASSO models, which have distinctive patterns in each province and thus help explain the city GDP variance across three provinces.

In addition, it is worth noting that if we took the province information as a dummy variable in the non-regularized MLR model of predicting GDP values using aforementioned traffic variables for all cities, the goodness of fit (R-squared) in the MLR model increased from 0.587 to 0.818, and the coefficients of this dummy variable had p-values close to 0. These values indicated that each province has its own structure of variability and such effect is statistically significant when predicting the GDP value from the transportation flow volumes of cars, buses, and trucks. 

In sum, this study demonstrates that highway transportation big data could reveal the status of regional economic development and contain valuable information of human mobility, production linkages, and logistics for regional management and planning. 

\section*{Methods} \label{section:methods}
\subsection*{Data} \label{section:data}
Annual transportation data between years 2014 and 2017 for three provinces, namely Shaanxi (10 cities), Jiangsu (13 cities) and Liaoning (14 cities), in China were collected and summarized in the city level. All vehicles are required to pay for their trips within the highway in China (Except in Hainan province). Using manual and electronic toll collection systems including radio-frequency identification (RFID) and automatic plate recognition technology as well as surveys in each station, detailed data including the entrance and exit stations, and the type of each vehicle (two categories: cars $\&$ buses, and freight trucks) were recorded. In addition, for cars and buses, the average number of passengers of each vehicle was estimated based on the type of cars and buses and the survey per vehicle type. Whereas for trucks, the total weight of each vehicle was also recorded and the loading weight was estimated based on the type of truck. More details about the data collection in highway toll stations in China can be found in previous works \cite{xiao2015trend,zhao2018analysis}. 
City level data were aggregated and summarized from the raw station-level data as our main focus was to investigate the relationship between regional economy and transportation networks.
We eliminated the traffic flows between different provinces and only kept the traffic flows within one province.
In sum, the total number of vehicles, the sum of passengers (for cars and buses) and the weights (for trucks) as well as the distance (km) between paired cities were calculated respectively (see Table \ref{t:data}).
In addition, the gross domestic product (GDP) data for each city in the same study period (2014-2017) were collected from the City Statistics Yearbook in each province, which measured the market value of all the final goods and services produced by all resident and institutional units engaged in a city.

\begin{table}[h]
\centering
	\scalebox{0.75}{
	\begin{tabular}{|c|c|c|c|c|c|}
	\hline
	Province                 & Number of Entry and Exit Stations & Year                  & Vehicle Type  & Number of Vehicles (million) & Number of Passengers or Volume of Weights (million) \\ \hline
	\multirow{8}{*}{Jiangsu}     & \multirow{8}{*}{421}& \multirow{2}{*}{2014}& Cars \& Buses  & 267.9  & 929.0 \\ \cline{4-6} 
	       & & & Trucks &     116.3 & 1150. 9 \\ \cline{3-6} 
	       & & \multirow{2}{*}{2015}& Cars \& Buses   &     259.8 & 639. 2 \\ \cline{4-6} 
	       & & & Trucks &     100.5 & 940,603.3 \\ \cline{3-6} 
	       & & \multirow{2}{*}{2016}& Cars \& Buses   &     422. 9 & 1289.4\\ \cline{4-6} 
	       & & & Trucks &     122.7 & 1,189,972.7 \\ \cline{3-6} 
	       & & \multirow{2}{*}{2017}& Cars \& Buses   &     486.5 & 1429.6 \\ \cline{4-6} 
	       & & & Trucks &     141.3 &  1,421,325.6 \\ \hline
	\multirow{8}{*}{Liaoning}     & \multirow{8}{*}{287}& \multirow{2}{*}{2014}& Cars \& Buses   &     107.0 & 398.1 \\ \cline{4-6} 
	       & & & Trucks &      41.1 & 429,189.2 \\ \cline{3-6} 
	       & & \multirow{2}{*}{2015}& Cars \& Buses   &     111.4 & 397.9 \\ \cline{4-6} 
	       & & & Trucks &      38.9 & 391,497.2 \\ \cline{3-6} 
	       & & \multirow{2}{*}{2016}& Cars \& Buses   &     121.0 & 424.9 \\ \cline{4-6} 
	       & & & Trucks &      43.4 & 445,433.7 \\ \cline{3-6} 
	       & & \multirow{2}{*}{2017}& Cars \& Buses   &     131.4 &  459.9 \\ \cline{4-6} 
	       & & & Trucks &      48.6 & 516,205.0 \\ \hline
	\multirow{8}{*}{Shaanxi}     & \multirow{8}{*}{335}& \multirow{2}{*}{2014}& Cars \& Buses   &     146.9 & 525.9 \\ \cline{4-6} 
	       & & & Trucks &      56.7 & 839,483.4 \\ \cline{3-6} 
	       & & \multirow{2}{*}{2015}& Cars \& Buses   &     165.3 & 566.1 \\ \cline{4-6} 
	       & & & Trucks &      57.3 &  787,819.5 \\ \cline{3-6} 
	       & & \multirow{2}{*}{2016}& Cars \& Buses   &     221.5 & 730.2 \\ \cline{4-6} 
	       & & & Trucks &      65.6 &     826,738.1\\ \cline{3-6} 
	       & & \multirow{2}{*}{2017}& Cars \& Buses   &     233.3 & 757.2 \\ \cline{4-6} 
	       & & & Trucks &      68.3 &     824,578.3 \\ \hline
	\end{tabular}}
	\caption{Summarized highway transportation data between 2014 and 2017 in three provinces: Jiangsu, Liaoning, and Shaanxi.}
	\label{t:data}
\end{table}

\subsection*{Estimating GDP based on the traffic flow data}
The multiple linear regression model (MLR) with ordinary least squares (OLS) parameter estimation is used to discover the statistical relationship between the transportation flow data and the economic development indicator (i.e. GDP). The GDP of a given city $i$ is considered as a dependent variable $GDP_i$ predicted by eight independent variables (also known as features or predictors) related to its transport flows of people and goods: intercity incoming flow of cars $\&$ buses ($I_{C}$), intercity outgoing flow of cars $\&$ buses ($O_{C}$), intracity flow of cars $\&$ buses ($N_{C}$), and the ratio of incoming/outgoing intercity flow for cars $\&$ buses ($R_{C}$), intercity incoming flow of trucks ($I_{K}$), intercity outgoing flow of trucks ($O_{K}$), intracity flow of trucks ($N_{K}$), and the ratio of incoming/outgoing intercity flow for trucks ($R_{K}$). The formula is as follows: 
\begin{equation}
GDP_i = b_0+b_1I_{Ci} + b_2O_{Ci}+ b_3N_{Ci} + b_4R_{Ci}+b_5I_{Ki} + b_6O_{Ki}+ b_7N_{Ki} + b_8R_{Ki} +e_i
\end{equation}

The intercity incoming flow ($I$) is the traffic flow with all origins outside the city and all destinations inside the city. The intercity outgoing flow ($O$) is the flow that originates inside the city and goes outside of the city. The intracity flow ($N$) is the flow with all origins and destinations inside the same city. The index ($R$) is the ratio of incoming and outgoing flows for a city, which may be able to indicate the role of a city compared with other cities among a transportation network. To demonstrate, when the ratio is larger than 1, it means there are more transportation flows coming to the city than leaving it, and the city may work as a `sink' in the network. It may be a commercial center attracting people coming to find jobs or an industry center consuming a large amount of goods. When the ratio is less than 1, it means that the city works more like a `source' which sends people or provides materials to other places. In addition, this study takes two types of traffic flows into consideration: the cars $\&$ buses flow and the truck flow. These two kinds of flows can reflect different aspects of the city economy, from the perspectives of human movement and goods movement.  There are in total eight independent variables together with the constant ($b_0$) and the error term ($e$) included in the multi-linear regression model. By solving this linear model using the OLS technique, the relationship between GDP and the traffic flow of each city in each year can be discovered. The model is first used for each province separately, in attempt to distinguish the pattern of economic development in each province. Then the data of three provinces are merged together to identify the general relationship of GDP and traffic flow across provinces. 

In addition, the standardized coefficient $b_j'$ of an independent variable from the multi-linear regression model is calculated as follows. It can be interpreted as the change of dependent variable in standard deviations, per standard deviation change in the predictors.

\begin{equation}
b_j'= b_j * \frac{S_{X_j}}{S_y}   
\end{equation}

Where $b_j$ is a regression coefficient. $S_y$ is the standard deviation of the dependent variable and $S_{X_j}$ is the standard deviation of independent variable $X_j$.

Besides, we also investigated a generalized linear model (GLM) with the natural log transformation of each city GDP given its log-normal or gamma distribution characteristics (in Supplement Fig. \ref{fig:GDP_hist_gamma_ggplot}). 
\begin{equation}
Ln(GDP_i) = b_0+b_1I_{Ci} + b_2O_{Ci}+ b_3N_{Ci} + b_4R_{Ci}+b_5I_{Ki} + b_6O_{Ki}+ b_7N_{Ki} + b_8R_{Ki} +e_i
\end{equation}

\subsection*{Linear models with regularization}
When some independent variables are highly correlated in the MLR, the OLS coefficient estimations will have a large variance. The introduced regularization techniques in statistics and machine learning help reduce variance at the cost of introducing some bias in a model and avoid the overfitting issue \cite{bickel2006regularization,scholkopf2001learning}. There are two types of regularization techniques with two different cost functions regarding the model complexity: the $L1$ norm (least absolute deviations) and the $L2$ norm (least squares). 

The Ridge regression applies the $L2$ penalty term to control the coefficient of each independent variable in a linear regression model \cite{hoerl1970ridge}. The objective of Ridge approach is to minimize the following cost function:
\begin{equation}
\sum_{i=1}^{N}(y_i-b_0-\sum_{j=1}^{p}x_{ij}b_{j})^2 + \lambda\sum_{j=1}^{p}b^2_{j}
\end{equation}
Where $y_i$ is the observed value of the dependent variable for each sample $i$ (i.e., the city GDP); $N$ is the total number of sample observations; $b_0$ is the constant in MLR;  $x_{ij}$ is the value of the independent variable $j$ for each sample $i$; $b_{j}$ is the estimated coefficient for the independent variable $j$ (abovementioned transportation flow variables); $p$ is the total number of independent variables (predictors),  and $\lambda >=0$ is a regularization penalty parameter. If $\lambda$ is 0, the cost function backs to the OLS estimation. However, if $\lambda$ is very large then it will add too much penalty to the model complexity and may lead to the model underfitting. The parameter calibration is performed to find the best $\lambda$ that fits the data well and achieves the bias-variance balance.

The LASSO regression has a similar cost function to minimize but it applies the $L1$ penalty term rather than the $L2$ norm to regularize the coefficient of each independent variable \cite{tibshirani1996regression}:
\begin{equation}
\sum_{i=1}^{N}(y_i-b_0-\sum_{j=1}^{p}x_{ij}b_{j})^2 + \lambda\sum_{j=1}^{p}|b_{j}|
\end{equation}

By shrinking some coefficients, the Ridge regression and the LASSO regression are able to control the multicollinearity in the model. It tends to pick one predictor from a few very correlated predictors and set the coefficients of the others to zero \cite{tibshirani1996regression, dong2019predicting}. Therefore, the regularization techniques are very helpful when there are many intercorrelated features and feature selection is necessary \cite{LinearRidgeLasso}.

\subsection*{The gravity model fitting with linear regression and linear programming}
To further estimate the relative attractions of cities and identify the important cities in each province, the gravity model is applied to the study area. This widely used model assumes that the flows between nodes are generated by some attraction force from the nodes\cite{okelly1995new}. With known flow values between cities, the attraction value of each city in this model can be estimated through a reverse calibration process \cite{okelly1995new}. 

The traffic flow $ {G_{ij}} $ is proposed to be computed by the following equation:
\begin{equation}
G_{ij} = k(P_iP_j)/(d_{ij})^\beta \;for\; all\;  i \neq j
\end{equation}

where \textit{k} is a constant and $d_{ij}$ is the distance between two cities. The attraction of each city($P_i$ and $P_j$) and the exponent on distance $\beta$ (i.e., the coefficient of distance decay effect) are unknown and need to be estimated. The equation can then be transferred by taking a natural logarithm.
\begin{equation}
lnG_{ij} = lnP_i+lnP_j-\beta(lnd_{ij}) + ln k
\end{equation}

By representing the formula with simpler symbols, we got the following equation:
\begin{equation}
b_{ij} + D_{ij} = X_i+X_j-a_{ij}\beta 
\end{equation}
where $X_i = lnP_i$,  $X_j = lnP_j$, $b_{ij} = lnG_{ij} - ln k $, $a_{ij} =lnd_{ij} $, $D_{ij}$ is the deviation between the estimation and the observation (i.e., the error term). 

Here the $b_{ij}$ can be considered as a known dependent variable and $X_i$ and $X_j$ are the coefficients of a series of dummy variables showing the relationship between cities \cite{okelly1995new}. For each given flow, the dummy variable will be 1 for the origin city and the destination city, and 0 for other cities. By forming such a linear regression relationship, it is possible to find the best
$X_i$, $X_j$, and $\beta$ that are closest to the real situation.

Besides the linear regression approach, a linear programming approach is also conducted as a comparison. It solves the problem starting from the deviation.
\begin{equation}
D_{ij} = X_i+X_j-a_{ij}\beta -b_{ij}  \; for\; all\;  i \neq j
\end{equation}

and $D_{ij}$ can be represented as 
\begin{equation}
D_{ij} = D_{ij}^{1} - D_{ij}^{2}
\end{equation}

$D_{ij}^{1}$ and $D_{ij}^{2}$ is always greater than or equal to 0. When $D_{ij}$ >0, $D_{ij}^{1}$ = $D_{ij}$ and $D_{ij}^{2}$ = 0;
When $D_{ij}$ <0, $D_{ij}^{1}$ = 0 and $D_{ij}^{2}$ = $-D_{ij}$. When $D_{ij}$  = 0, both $D_{ij}^{1}$ and  $D_{ij}^{2}$ is 0 \cite{ecker2004introduction}.
The goal of the linear programming is to minimize the maximum absolute error (MINIMAX). Here the  $M$ represents the maximal absolute deviation and the optimization model is listed as:
\begin{flalign*}
Minimize :  M \\
Subject \;  to: \\
\end{flalign*}
\begin{equation}
D_{ij}^{1} - D_{ij}^{2} = X_i+X_j-a_{ij}\beta -b_{ij}
\end{equation}
\begin{equation}
M - D_{ij}^{1} - D_{ij}^{2} >= 0
\end{equation}
\begin{equation}
D_{ij}^{1}, D_{ij}^{2},X_i,X_j,\beta, M >=0
\end{equation}

By solving this model, the values of $X_i$ for all cities and the distance-decay coefficient $\beta$ can be estimated. The attraction of each city and the $\beta$ values of four years calculated from the MINIMAX approach and the linear regression approached are then compared.

\subsection*{The null model}
A null model is also used to examine the effect of distance decay (i.e. the value of $\beta$). The null model first constructs an interactive network without considering the distance among nodes\cite{liu2014quantifying,zhuo37distance}. Then by comparing the estimated flow in the null model with the observed flow in reality, the difference can reflect the effect of distance. The total flow of a specific node can be denoted as:
\begin{equation}
W_i = \sum_jG_{ij}
\end{equation}
Letting $ F=\sum_{i}\sum_{j} G_{ij}$ and $N=\sum_{i}\sum_{j} W_iW_j $, and the estimated flow between node $i$ and $j$ is:
\begin{equation}
G_{ij}^{null} = W_iW_jF/N
\end{equation}
The ratio $ R_{ij} = G_{ij}/G_{ij}^{null} $ is used to measure the effect of distance decay and since it should be related to the distance itself. A linear regression model between $ R_{ij} $ and $d_{ij}$ is constructed. Then the best fitted slope is taken as the value of the distance-decay coefficient $\beta$. 

\subsection*{Network structure analyses}
The centrality and PageRank measures are often used in transportation network analysis to quantify the importance of a node in road networks \cite{chen2012identifying,gao2013understanding,zhao2017network}. Given a network $G<V,E,W>$ of nodes (V) and edges (E) with weights (W),  The betweenness and closeness centrality measures are defined as follows. The shortest paths can be calculated based on different weights such as physical distance and traffic volume, as discussed in the section "Network measures and sub-network structure". 

\par The betweenness centrality of a node $Betw(i)$ quantifies how often the shortest paths passes through a node $i$ in a network.  
\begin{equation}
Betw(i) = \sum_{j \neq k \neq i}\frac{n_{jk}(i)}{n_{jk}}
\end{equation}
Where $n_{jk}$ is the number of shortest paths between nodes $j$ and $k$, and $n_{jk}(i)$ is the number of shortest paths between nodes $j$ and $k$ that contain node $i$.

\par The closeness centrality of a node $Clos(i)$ is the reciprocal of the sum length of the shortest paths between the node and all other nodes in the network.
\begin{equation}
Clos(i) = \frac{1}{\sum^n_j d_{ij}}
\end{equation}
Where $n$ is the total number of nodes; $d_{ij}$ is the length of the shortest path between nodes $i$ and $j$.

\par The PageRank $PR(i)$ quantifying the node importance in networks is computed as follows \cite{brin1998anatomy,mao2015quantifying}. 
\begin{equation}
PR(i) = \frac{1-d}{n} + d*\sum_{j\in M(i)} PR(j) * W_{ij}/L(j) 
\end{equation}
Where $d$ is a damping factor between 0 and 1 and is usually set as 0.85 \cite{brin1998anatomy}; $n$ is the total number of nodes; $j$ is the index of a set of nodes $M(i)$ that link to $i$, $L(j)$ is the number of outbound links on node $j$, and $PR(j)$ is the PageRank of node $j$. As for the weighted networks, the PageRank algorithm interprets the normalized edge weight $W_{ij}$ as connection strengths and an edge with a larger weight is more likely to be selected for connection.

In addition, the Principle Component Analysis (PCA) is utilized to identify the structure of the network and discover potential sub-networks inside a province \cite{demvsar2013principal}. By applying a standard PCA to the transport flow matrix $[F_{ij}]$, every principal component (PC) can be related to a sub-system of flows and the first few components represent stronger sub-systems \cite{demvsar2013principal,goddard1970functional}. The factor loadings and component scores are computed for each PC and they are used to identify the dominant sub-networks. Given a dimension of $n * m$ flow data set $F_{ij}$, the first principal component of a set of destination features $X_1$,$X_2$, ..., $X_j$, $X_m$,  is the normalized linear combination of the features that has the largest variance. As shown below, the elements $\phi_{11},…,\phi_{m1} $ as the factor loadings of the first principal component. The loading vector defines a direction in feature space along which the data varies most. Similarly, we can derive other at most $min(n-1, m)$ principle components.

\[\begin{bmatrix}
F_{11}&F_{12}&...&F_{1j}&...F_{1m}\\
F_{21}&F_{22}&...&F_{2j}&...F_{2m}\\
...\\
F_{i1}&F_{i2}&...&F_{ij}&...F_{im}\\
...\\
F_{n1}&F_{n2}&...&F_{nj}&...F_{nm}\\
\end{bmatrix}\]

\begin{equation}
	Z_{1} = \phi_{11}X_{1} + \phi_{21}X_{2}  + . . . + \phi_{j1}X_{j}   + . . .+ \phi_{m1}X_{m}
\end{equation}

\begin{equation}
X_{j} = (F_{1j}, F_{2j}, ..., F_{ij}, ... , F_{nj} )   
\end{equation}


\bibliography{references}

\section*{Acknowledgments}
Bin Li and Runmou Xiao would like to thank the funding support of this research by the Fundamental Research Funds for the Central Universities (Grant No. 300102229108), CHD; Transportation Strategic Planning Policy Project Plan 2018 (Grant No. 2018-22-3), Ministry of Transport of the People's Republic of China. Song Gao would like to thank the research support by the University of Wisconsin - Madison Office of the Vice Chancellor for Research and Graduate Education with funding from the Wisconsin Alumni Research Foundation.

\section*{Author contributions statement}
B.L., S.G., and R.M.X. designed this research; B.L., S.G., Y.L.L. and  Y.H.K. performed experiments, analyzed data and wrote the paper; T.P. created the maps and data visualization; Y.Q.G wrote the paper.  All authors reviewed and edited the manuscript. 

\section*{Additional information}
Supplementary information accompanies this paper at http://www.nature.com/scientificreports
Competing financial interests: The authors declare no competing financial interests.
How to cite this article:.XXX

\pagebreak
\section*{Appendix} \label{section:supplementary}

\beginsupplement
\subsection*{Modeling the GDP distribution}
\begin{figure}[H]
	\centering
	\includegraphics[width=0.8\linewidth]{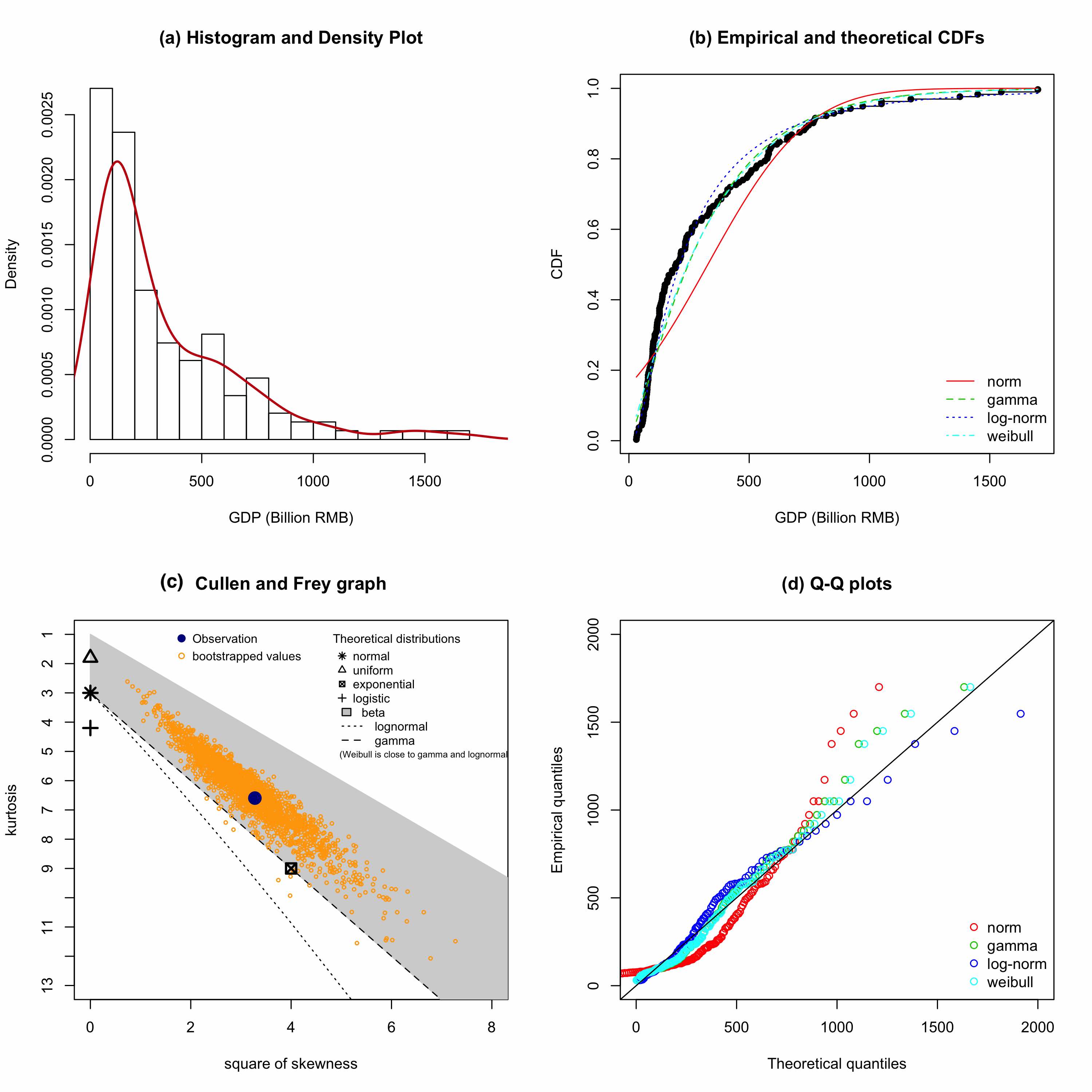}
	\caption{The empirical and theoretical distributions of all city GDP values. (a) the histogram of city GDPs; (b) the observed empirical cumulative distribution function (CDF) in black and four  theoretical CDFs; (c) the Cullen and Frey graph shows the probability distribution fittings; (d) the Q-Q plot compares the empirical and theoretical quantile distributions.}
	\label{fig:GDP_hist_gamma_ggplot}
\end{figure}

The empirical histogram distribution of the economic development indicator (GDP) values across all cities in the study period is a positively skewed distribution with a long-tail end (Fig. \ref{fig:GDP_hist_gamma_ggplot}a). By fitting the data observations into four theoretical models (i.e., normal, gamma, log-normal, and Weibull distributions) using the maximum likelihood estimation \cite{delignette2015fitdistrplus}, the results show that the log-normal model is the best with the maximum log-likelihood of -994.39 and the minimum Akaike information criterion (AIC) value of 1992.78. AIC is used to determine which model performs the best regarding the goodness of fit and the simplicity of a model using information theory \cite{akaike1998information}. Meanwhile, the log-likelihood and AIC values for other models are  gamma distribution (log-likelihood=-1002.82 $\&$ AIC=2009.64), Weibull distribution (log-likelihood=-1004.23 $\&$ AIC=2012.45), and normal distribution (log-likelihood=-1066.17 $\&$ AIC=2136.35) respectively. Fig. \ref{fig:GDP_hist_gamma_ggplot}b and Fig. \ref{fig:GDP_hist_gamma_ggplot}d show the curves of empirical and theoretical cumulative distribution functions of these models and their quantile-quantile plots. The skewness-kurtosis plot (i.e., the Cullen and Frey graph \cite{cullen1999probabilistic}) is used to examine the characteristics of the skewness and the degree of tailedness (i.e., kurtosis) of the GDP data with uncertainty compared with the theoretical models. As shown in the \ref{fig:GDP_hist_gamma_ggplot}c, the bootstrapping values fit better with the log-normal, gamma and Weibull models rather than the normal, uniform and logistic models. The consistent positive skewness and kurtosis values verify the "heavy-tailed on the right" characteristic of the city GDP distribution in the study areas. 

\begin{figure}[!htb]
	\begin{subfigure}[b]{0.4\linewidth}
		\includegraphics[width=\linewidth]{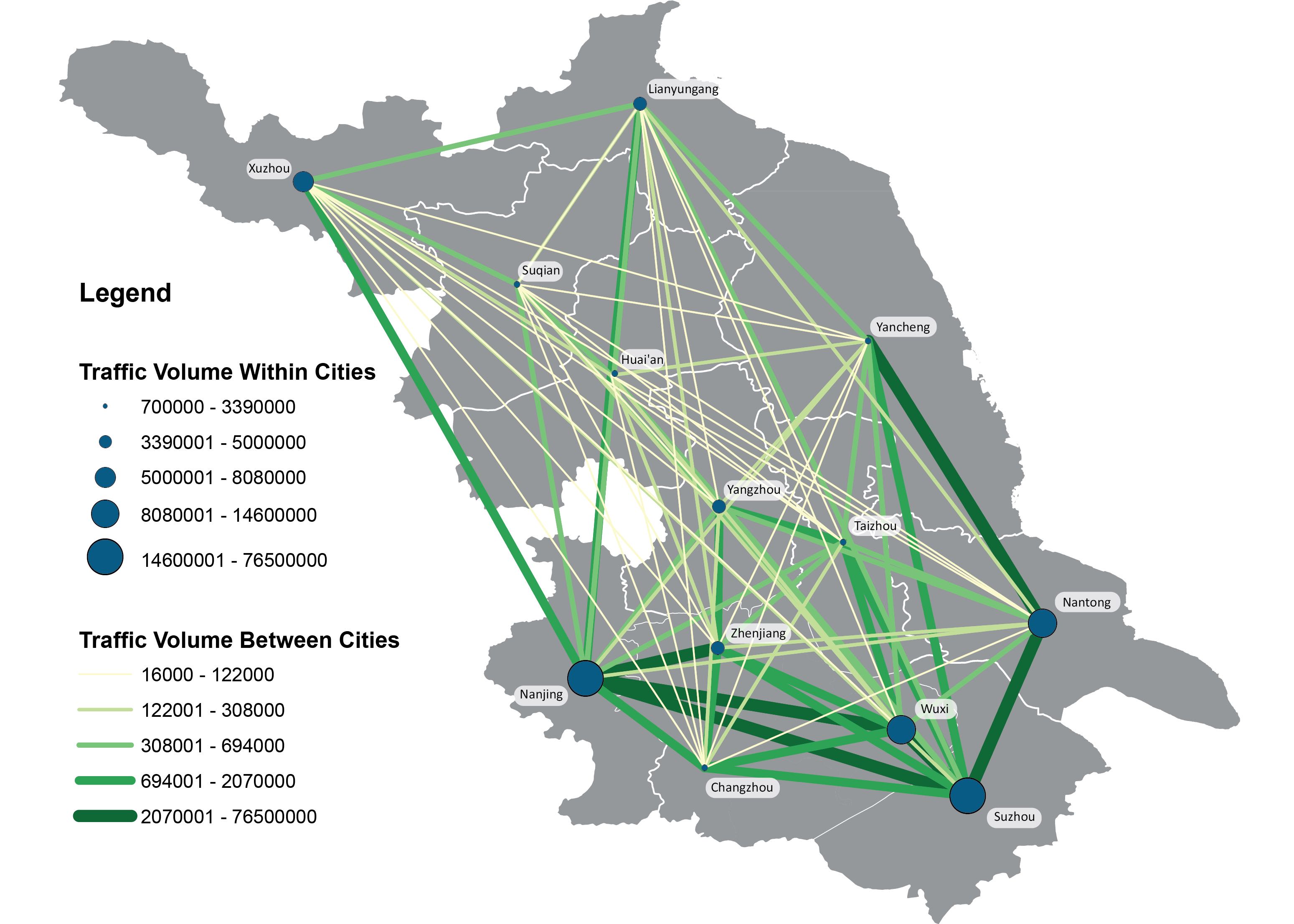}
		\caption{2014}
	\end{subfigure}
	\begin{subfigure}[b]{0.4\linewidth}
		\includegraphics[width=\linewidth]{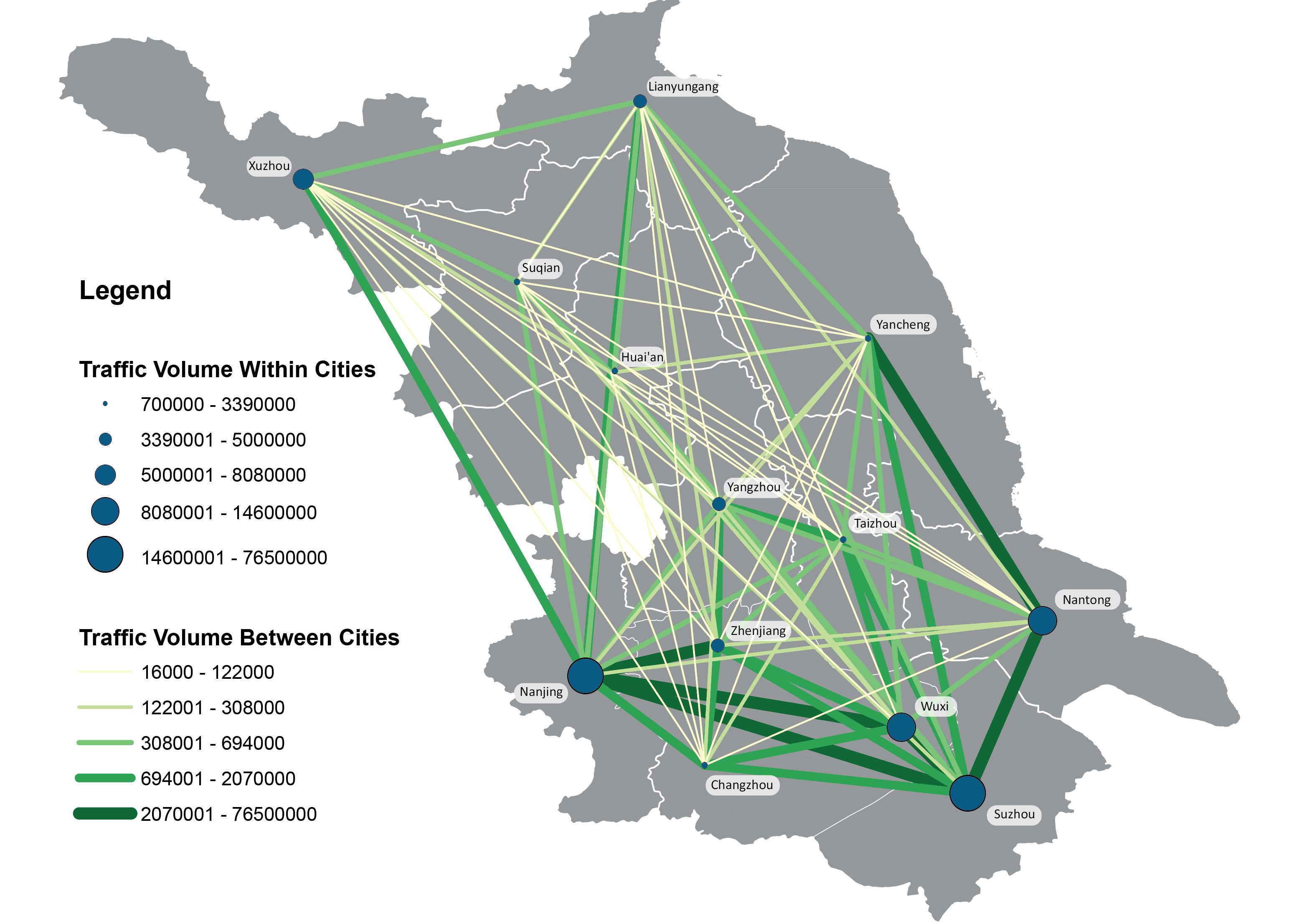}
		\caption{2015}
	\end{subfigure}
	\begin{subfigure}[b]{0.4\linewidth}
		\includegraphics[width=\linewidth]{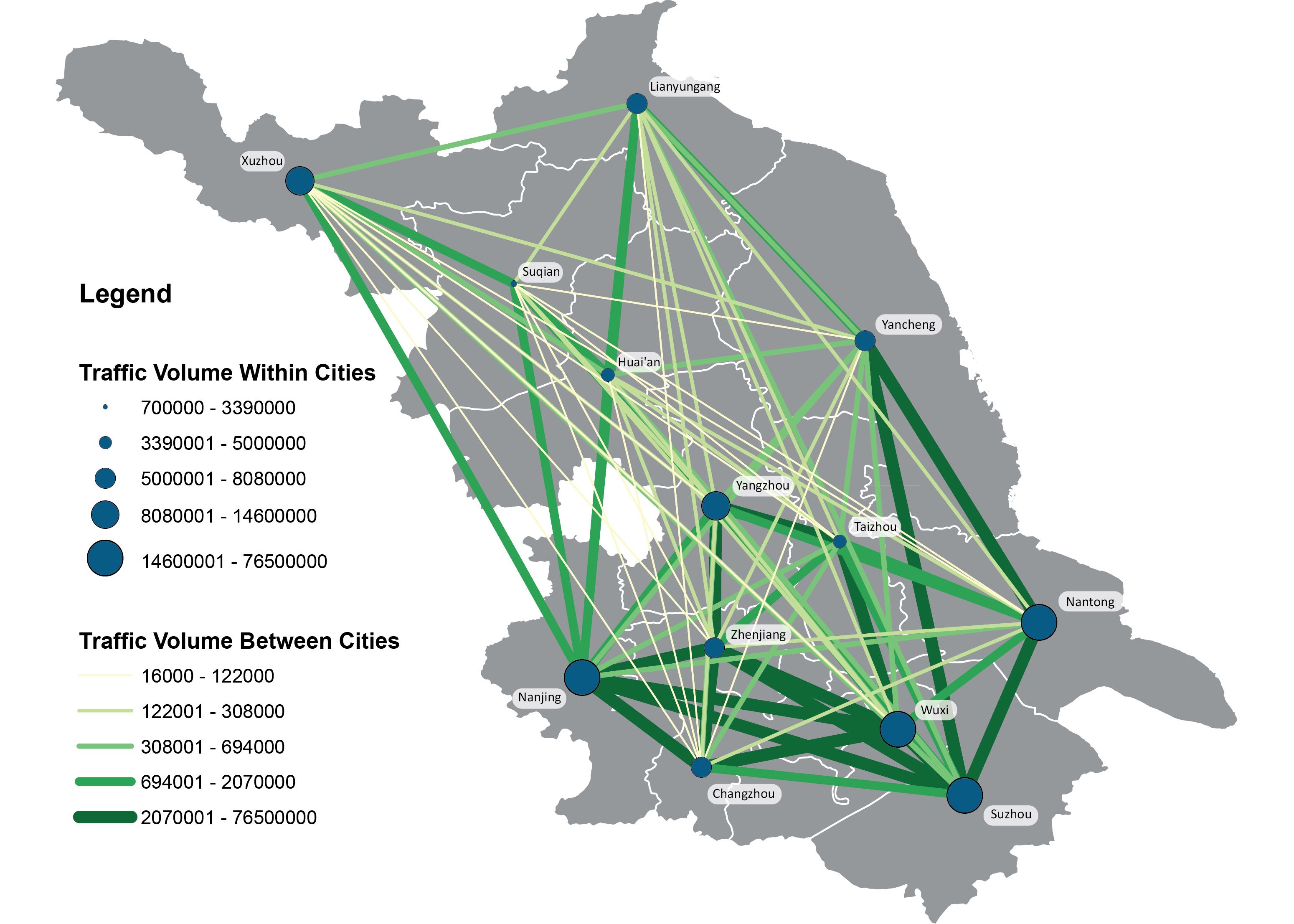}
		\caption{2016}
	\end{subfigure}
	\begin{subfigure}[b]{0.4\linewidth}
		\includegraphics[width=\linewidth]{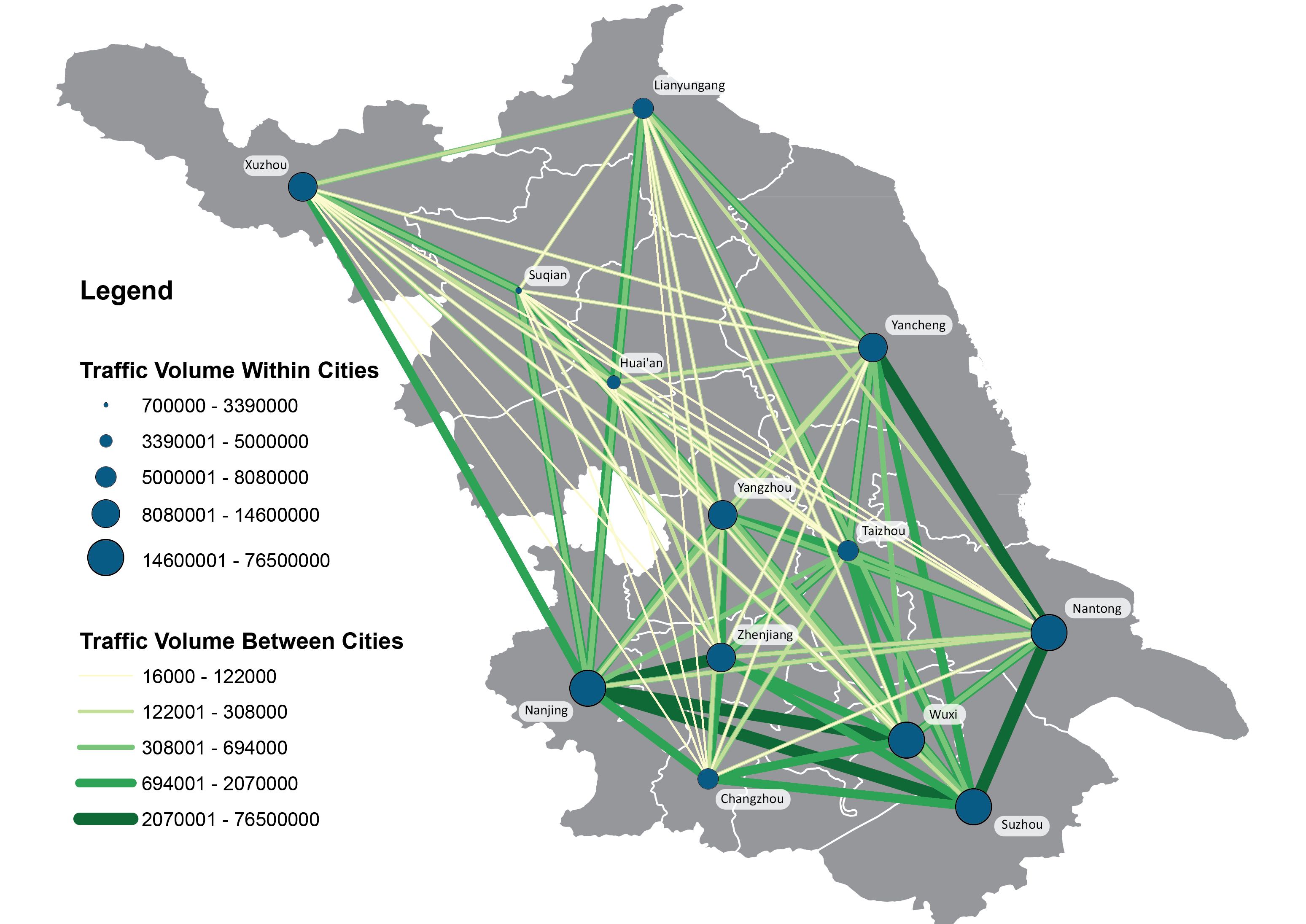}
		\caption{2017}
	\end{subfigure}
	\centering
	\caption{Traffic Volume of Cars and Buses in Jiangsu. Note: The maps were generated using ArcMap version 10.6 and Adobe Illustrator CC version 20.}
		\label{fig:JSflows}
\end{figure}

\begin{figure}[!htb]
	\begin{subfigure}[b]{0.4\linewidth}
		\includegraphics[width=\linewidth]{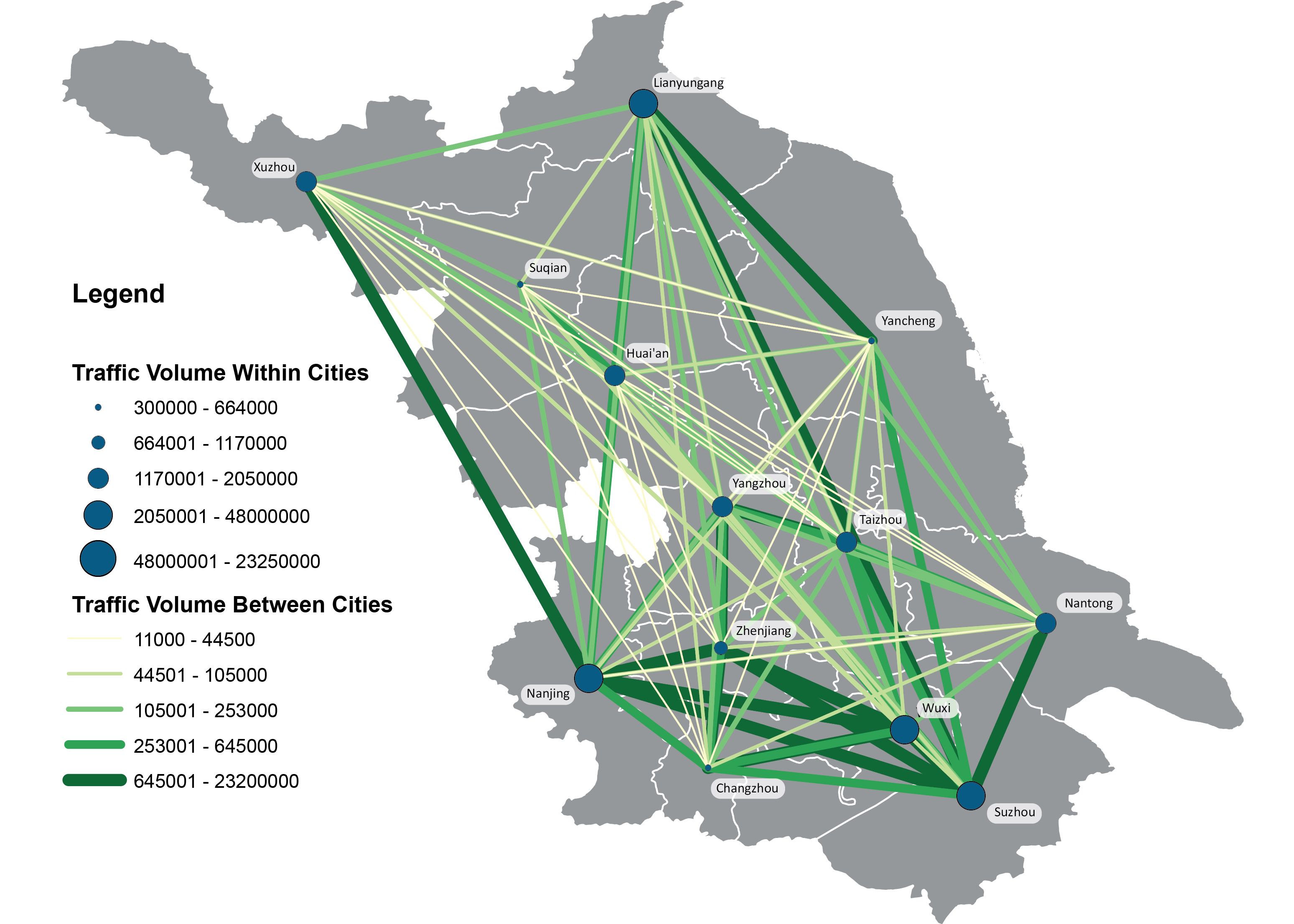}
		\caption{2014}
	\end{subfigure}
	\begin{subfigure}[b]{0.4\linewidth}
		\includegraphics[width=\linewidth]{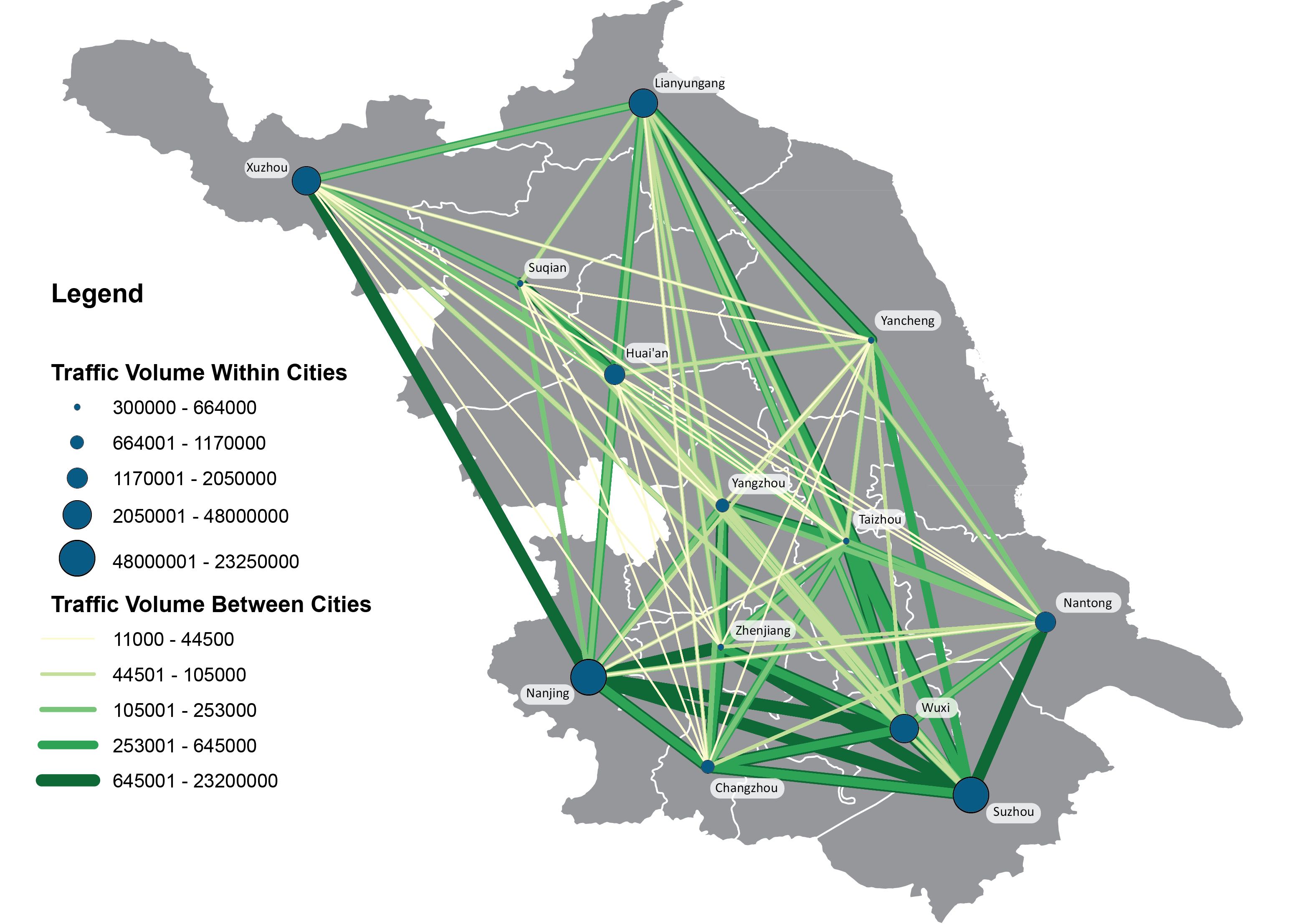}
		\caption{2015}
	\end{subfigure}
	\begin{subfigure}[b]{0.4\linewidth}
		\includegraphics[width=\linewidth]{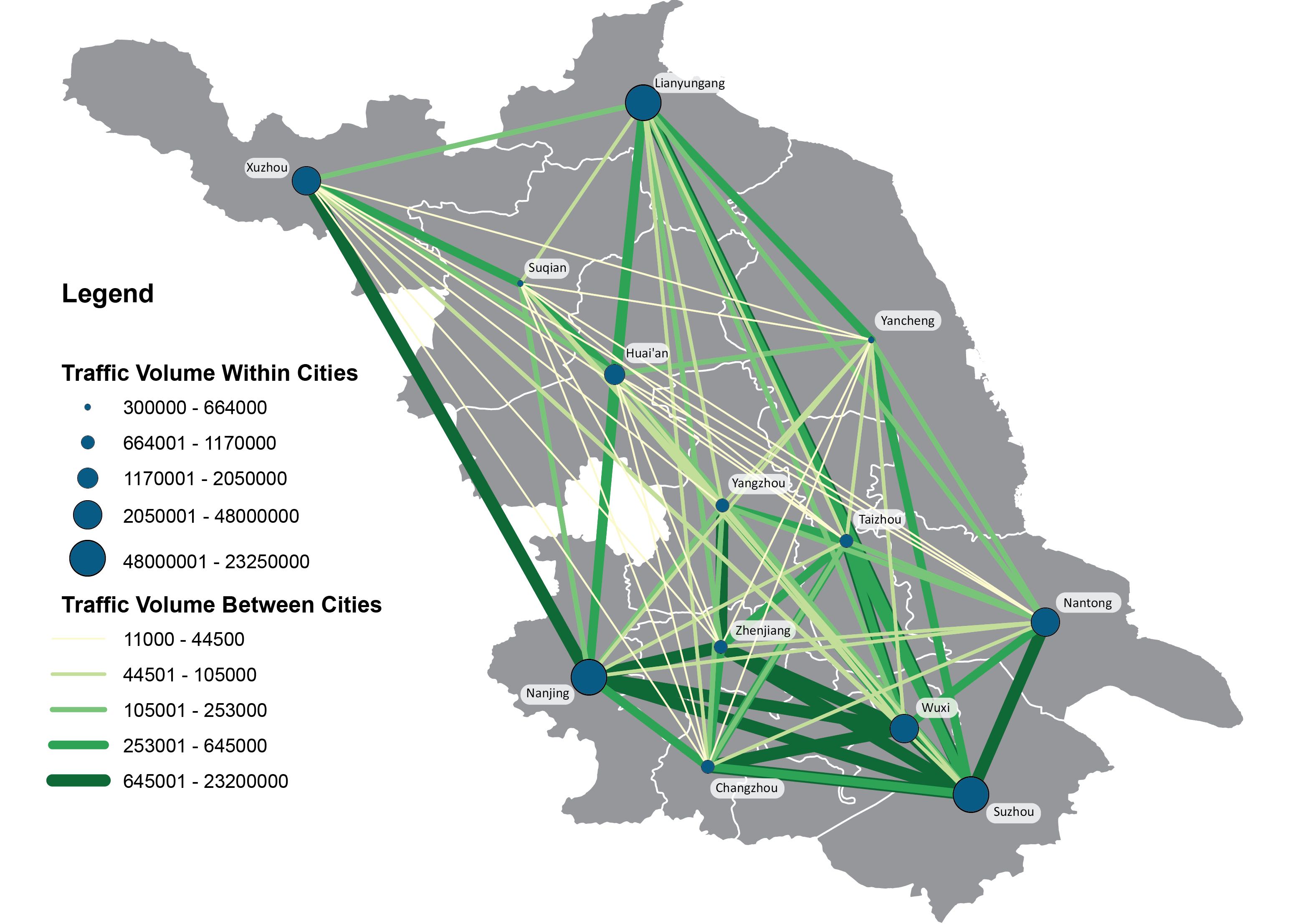}
		\caption{2016}
	\end{subfigure}
	\begin{subfigure}[b]{0.4\linewidth}
		\includegraphics[width=\linewidth]{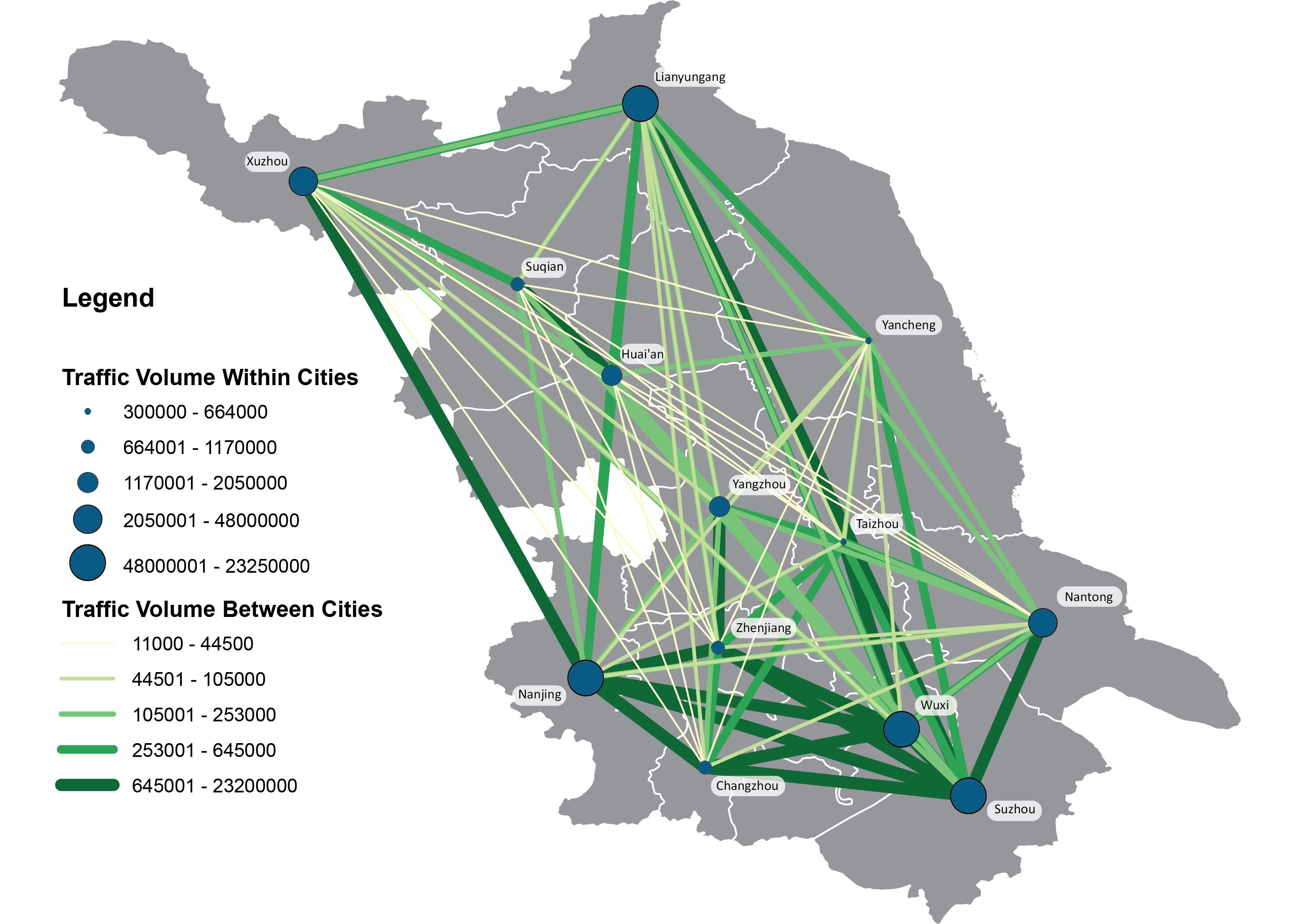}
		\caption{2017}
	\end{subfigure}
	\centering
	\caption{Traffic Volume of Freight Trucks in Jiangsu. Note: The maps were generated using ArcMap version 10.6 and Adobe Illustrator CC version 20.}
	\label{fig:JSflows_truck}
\end{figure}

\end{document}